\documentclass[aps,pra,letterpaper,10pt,twocolumn,notitlepage,superscriptaddress]{revtex4-1}

\usepackage{amsmath, amsfonts, times, mathtools, qcircuit, xcolor, bm, bbm, graphicx, hyperref, cleveref, tikz}
\usepackage[T1]{fontenc}
\usepackage{soul}
\hypersetup{
    colorlinks,
    linkcolor={blue},
    citecolor={blue},
    urlcolor={blue},
    pdfauthor={Robin Harper, Steven T Flammia, Joel J Wallman},
    pdftitle={Efficient learning of quantum noise}
}
\usetikzlibrary{external}
\usepackage{longtable}
\usepackage{amsmath, amsthm, amsfonts, amssymb, xcolor, bm, bbm, graphicx, hyperref, cleveref, tikz}
\usepackage[T1]{fontenc}
\usepackage[pass]{geometry}
\usepackage{soul}
\hypersetup{
    colorlinks,
    linkcolor={blue},
    citecolor={blue},
    urlcolor={blue},
    pdfauthor={Robin Harper, Steven T Flammia, Joel J Wallman},
    pdftitle={Efficient learning of quantum noise}
}
\usepackage{etoolbox} 
\usetikzlibrary{external}
\graphicspath{{figures/}}

\theoremstyle{plain}

\theoremstyle{definition}

\newcommand{\bs}[1]{\boldsymbol{#1}}

\begin{document}

\title{Efficient learning of quantum noise}

\author{Robin Harper}
\affiliation{Centre for Engineered Quantum Systems, School of Physics, University of Sydney, Sydney, NSW 2006 Australia}
\author{Steven T. Flammia}
\thanks{Corresponding author: sflammia@gmail.com}
\affiliation{Centre for Engineered Quantum Systems, School of Physics, University of Sydney, Sydney, NSW 2006 Australia}
\affiliation{Yale Quantum Institute, Yale University, New Haven, CT 06520, USA}
\author{Joel J. Wallman}
\affiliation{Institute for Quantum Computing and Department of Applied
Mathematics, University of Waterloo, Waterloo, Ontario N2L 3G1, Canada}
\affiliation{Quantum Benchmark Inc., 100 Ahrens Street West, Suite 203, Kitchener, ON N2H 4C3, Canada}

\date{\today}

\begin{abstract}
{Noise is the central obstacle to building large-scale quantum computers.
Quantum systems with sufficiently  uncorrelated and weak noise could be used to solve computational problems that are intractable with current digital computers.
There has been substantial progress towards engineering such systems~\cite{Schindler2011, Reed2012, Barends2014, Nigg2014, Kelly2015, Corcoles2015, Ofek2016, Linke2017}.
However, continued progress depends on the ability to characterize quantum noise reliably and efficiently with high precision~\cite{Martinis2015}.
Here we describe such a protocol and report its experimental implementation on a 14-qubit superconducting quantum architecture. The method returns an estimate of the effective noise and can detect correlations within arbitrary sets of qubits. We show how to construct a quantum noise correlation matrix allowing the easy visualization of correlations between all pairs of qubits, enabling the discovery of long-range two-qubit correlations in the 14 qubit device that had not previously been detected. Our results are the first implementation of a
provably rigorous and comprehensive diagnostic protocol capable of being run on state of the art devices and beyond. These results pave the way for noise metrology in next-generation quantum devices, calibration in the presence of crosstalk, bespoke quantum error-correcting codes~\cite{Tuckett2018}, and customized fault-tolerance protocols~\cite{Aliferis2008} that can greatly reduce the overhead in a quantum computation.}
\end{abstract}
\maketitle

Useful large-scale quantum computers will require both careful calibration to reduce errors and some form of error correction before universal quantum computing can be realized.
Due to crosstalk, optimal calibrations of gates depend on the other gates that are being implemented, which can reduce the overall system error rate by an order of magnitude~\cite{Neill2018}.
Furthermore, error correction routines rely on knowing what the most likely error sources are.
Error correction routines that are optimized for the specific noise in a system can dramatically outperform generic ones~\cite{Aliferis2008, Tuckett2018}.

\begin{figure*}[ht!]
\includegraphics[width=\textwidth]{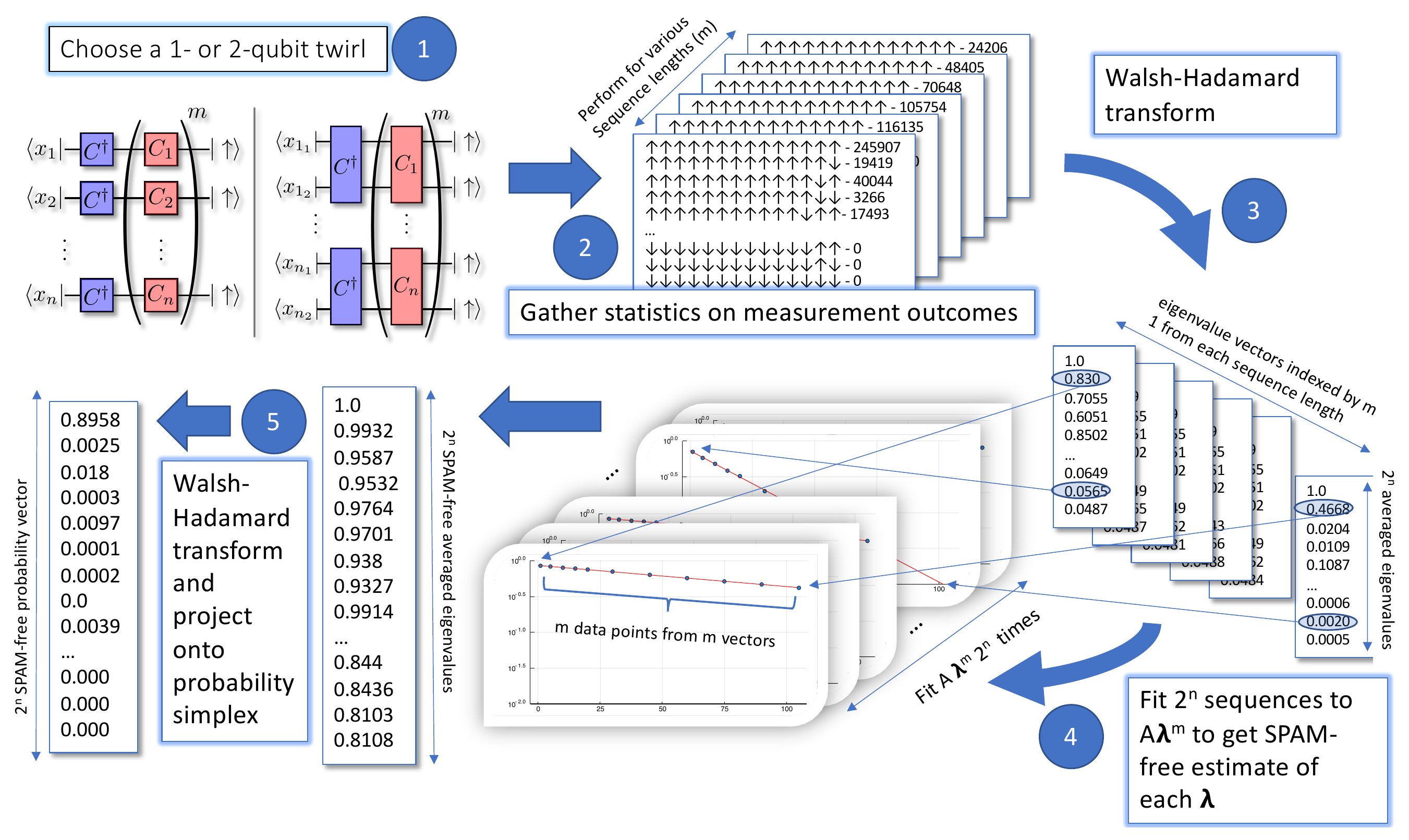}
\caption{\label{fig:protocol}
Our algorithm for characterizing the entire averaged probability vector. 
In step 1 we determine whether we wish to examine the correlations between qubits activated in single-qubit mode or when gates between the qubits are also used. 
Each red box in the quantum circuits of step 1 denotes a sequence of random Clifford gates, and the circuits themselves are run for varying lengths $m$ (with time moving from right to left).
The final gate $C^\dagger$ is not random, but is rather an inversion gate returning the system to its initial state in the ideal case.
Step 2 gathers output statistics from the measurement results $x_i$ obtained in the experiments in step 1, and the empirically estimated probability distributions for these outcomes are transformed in Step 3 via a Walsh-Hadamard transform. 
The transformed values are each fit to an exponential decay in Step 4, allowing us to reconstruct the averaged eigenvalues free of state preparation and measurement error (SPAM).
With a final reverse transformation in Step 5, we reconstruct the entire list of effective observed error rates.}
\end{figure*}

The calibration and error correction necessary for useful large-scale quantum computing therefore depends upon the ability to characterize the noise in large quantum systems. 
This characterization will become increasingly important as the field continues to progress~\cite{Martinis2015}. 
Unfortunately, current methods of characterizing noise are infeasible or report only a single summary statistic (e.g., the process fidelity) as device sizes approach 10 or more qubits, which is already the state of the art.
Process tomography~\cite{Chuang1997}, gate set tomography~\cite{Blume-Kohout2016}, and robust process tomography~\cite{Kimmel2013} do not scale past a handful of qubits even when sophisticated techniques such as compressed sensing~\cite{Gross2010, Flammia2012, Riofro2017} are utilized. 
Randomized benchmarking~\cite{Emerson2005, Knill2008} (RB) is a protocol that does scale in principle, but it provides only an incomplete description of the noise. 
Recent generalizations of RB improve its scalability and applicability~\cite{Franca2018, Proctor2018, Wallman2015a, Helsen2018, Erhard2019}.
However, these generalizations still aim to provide a single figure of merit summarizing the noise, which does not capture all the relevant information about how errors affect large-scale quantum devices.
Some of the techniques related to randomized benchmarking provide more tomographic information~\cite{Emerson2007, Gambetta2012, Erhard2019}.
However, the protocol of Ref.~\cite{Emerson2007} is not robust to SPAM errors while the protocols of Ref.~\cite{Gambetta2012, Erhard2019} lack the crucial digital processing steps needed to efficiently estimate correlations within arbitrary sets of qubits.

Here we develop and experimentally implement a protocol based on the general method of Ref.~\cite{Flammia2019} that allows us to learn a complete description of the observed error rates (see supplementary information for a full definition) in a large-scale quantum device. 
Where the device is too large to allow a complete description of these error rates to be written or sufficient data to be practically gathered  (say, $\gg 20$ qubits), we show how to model the system in such a way that, with only mild and physically plausible assumptions, we can reconstruct the effective observed error rates to arbitrary model fidelity. 
The protocol is efficient in $n$, the number of qubits, and comes with mathematically rigorous guarantees on its convergence and performance assuming only mild and physically plausible assumptions. 
Furthermore, the method is immune to systematic bias due to state preparation and measurement errors (SPAM), and achieves both high precision and accuracy.

For any given noisy quantum system comprising $n$ qubits, we can consider the average noise to have the special form of a Pauli channel~\cite{Knill2005}. 
Although not every noise channel is a Pauli channel, practical methodologies have been developed to transform the noise to be exceptionally well-approximated by a Pauli channel without introducing new errors~\cite{Wallman2016, Ware2018}. 
A Pauli channel $\mathcal{E}$ acting on a quantum state $\rho$ is of the form $\mathcal{E}(\rho) = \sum_j p(j) P_j \rho P_j$, where $p(j)$ is the error rate associated with the Pauli operator $P_j$. 
The Pauli error rates $p(j)$ form a probability distribution. 
These are closely related to, but distinct from, the eigenvalues of the Pauli channel (the \textit{eigenvalues}), which are defined to be $\lambda(j) = 2^{-n}\mathrm{Tr}\bigl(P_j \mathcal{E}(P_j)\bigr)$. 
Thus, when a state $\rho$ is subjected to the noisy channel $\mathcal{E}$, the Pauli error rate $p(j)$ describes the probability of a multi-qubit Pauli error $P_j$ affecting the system, while the respective eigenvalue describes how faithfully a given multi-spin Pauli operator is transmitted.
The Pauli error rates $p(j)$ and eigenvalues $\lambda(j)$ are related by a Walsh-Hadamard transform.
The (rescaled) average value of $\lambda(j)$, which is measured by RB, is the only figure of merit estimated and reported by most quantum computing experiments. A protocol to measure a symmetrized characterization of the noise to additive precision was presented by Emerson et al~\cite{Emerson2007}.

Obtaining a complete description of the Pauli error rates requires learning more than just the single-Pauli eigenvalues or single-Pauli error rates~\cite{Wright2019}, i.e., those associated with the single-qubit Pauli operators such as $\sigma_z^{(1)}$ or $\sigma_x^{(3)}$.
A complete description requires learning all of the noise correlations in the system, that is, how the probabilities of multi-qubit Pauli operators, e.g., $\sigma_z^{(1)}\otimes \mathbbm{1}_{\vphantom{x}}^{(2)}\otimes\sigma_x^{(3)}$, differ from the probabilities predicted under independent local noise.
Knowing these correlations is essential for removing unwanted correlated errors~\cite{Gambetta2012} and for performing optimal quantum error correction~\cite{Chubb2018}. 
The number of all possible noise correlations grows exponentially with the number of qubits, so it is crucial to have an efficient description of the most relevant correlations in order to focus error-mitigation efforts on the dominant noise sources.
The protocol and analysis presented here shows how to do this using noise that has been averaged over the local Pauli basis (the quantum noise), although the full protocol of Ref.~\cite{Flammia2019} can be used to learn the noise without averaging over the local Pauli basis. 
We make these terms more precise later.

\autoref{fig:protocol} gives a complete description of our protocol for learning quantum noise.
It proceeds in five steps: First, as in simultaneous randomized benchmarking~\cite{Gambetta2012}, choose sequences of random quantum gates of different lengths chosen independently from the single-qubit (or two-qubit) Clifford group, i.e., the group generated by the Hadamard, Phase, and (for two-qubits) CNOT gates on each individual qubit or qubit pair. 
Second, for each sequence length parameter $m$, estimate the resulting empirical probability distributions of the $n$-qubit measurement outcomes averaged over the sequences of length $m$.
Third, take a Walsh-Hadamard transform of each of the empirical probability distributions. 

The resulting values will exhibit an exponential decay with respect to the sequence length parameter, $m$, so in step 4 we fit to such a model to learn the decay constants. 
Finally, we transform back and project onto the probability simplex. 

This procedure provably converges to an estimate of the probability distribution of the average noise in the system~\cite{Flammia2019}, though the variant implemented here uses random gates chosen from the single-qubit Clifford group (as in simultaneous RB~\cite{Gambetta2012}) instead of the Pauli group. 
This leads to a simpler protocol, but one that additionally averages over the local basis information.
Averaging the noise in this way reduces the number of parameters that need to be reconstructed from $4^n$ to $2^n$.
We call this reduced distribution the observed error rates to contrast with the larger distribution of Pauli error rates, and we similarly define the set of $2^n$ averaged eigenvalues in analogy with the $4^n$ eigenvalues. 
Importantly, the observed error rates are still capable of describing many-body correlations within arbitrary sets of qubits. 

\begin{figure}[th!]
\includegraphics[width=0.95\columnwidth]{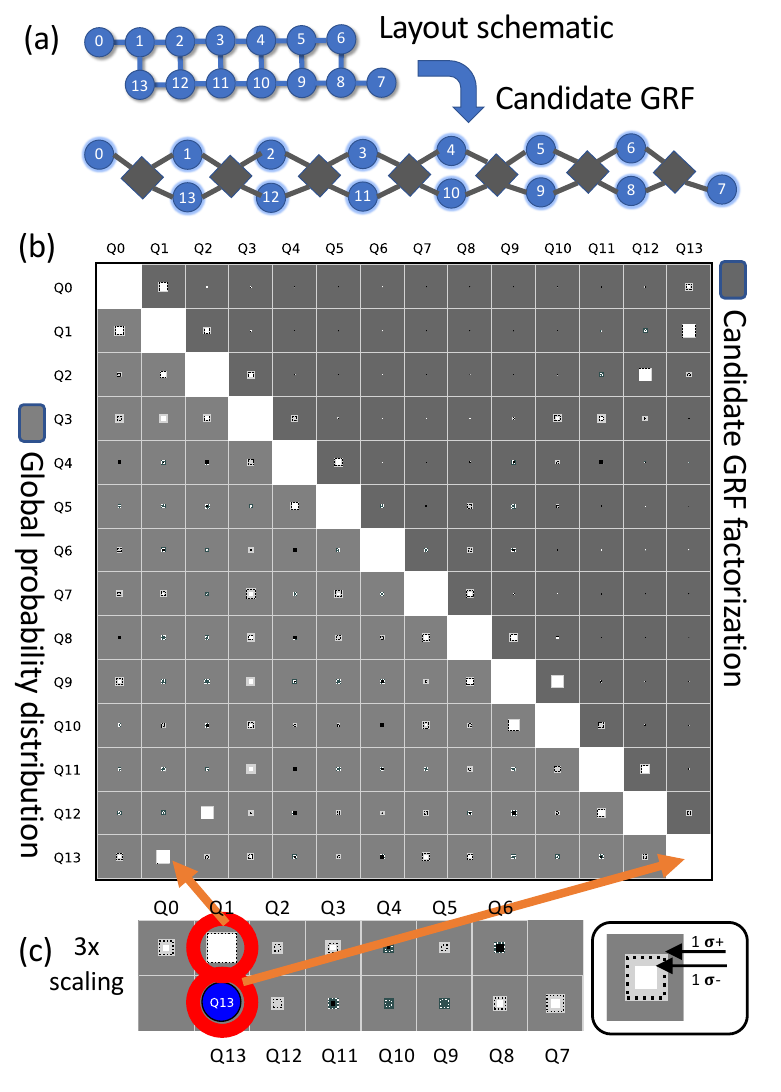}
\caption{\label{fig:singleCorrelation} Results from the protocol run in single-qubit mode.
(a) Spatial layout of the qubits in the 14 qubit \textit{Melbourne} architecture. 
Edges in the schematic graph correspond to qubit pairs that can be coupled via a two-qubit gate.
The factor graph below that is for a Gibbs random field (GRF) that models quantum noise via spatially local correlations in the device. 
Each diamond-shaped node is a factor that can describe arbitrary correlations among the qubits connected to it. 
(b) We can use the reconstructed probability distribution to look at the correlations between the probability of an error on each of the qubits, where the probability of an error on a particular qubit is treated as a random variable. These correlations can be plotted in the form of a correlation matrix calculated from the globally estimated distribution.
Correlation matrices are always symmetric, so we plot separately the correlations from the global estimate (lower left) and the GRF reconstruction (upper right) assuming the factor graph in (a). 
Gray background indicates a value of zero, and white (black) boxes indicate positive (negative) values between $[-1,1]$ in proportion to their relative area.
c) An example of how the spatial correlation of the qubits on the device translate to the layout in the correlation matrix. 
This example shows that although qubit 1 and qubit 13 are spatially adjacent, they are not adjacent in the matrix. 
To the right, we illustrate the convention used to indicate error bars, using here $1\sigma$ bounds.}
\end{figure}

The protocol presented in \autoref{fig:protocol} reconstructs the full probability distribution with a number of experiments that scales polynomially in the number of qubits $n$, but requires computational resources that scale with the (generally exponential) number of error rates to be estimated. 
The number of observed error rates scales as $2^n$, so to make our protocol truly scalable, we need a method to estimate an efficient description of the noise that nonetheless captures the correlations in a transparent, systematic, and physically motivated way.

\begin{figure}[t!]
\includegraphics[width=0.95\columnwidth]{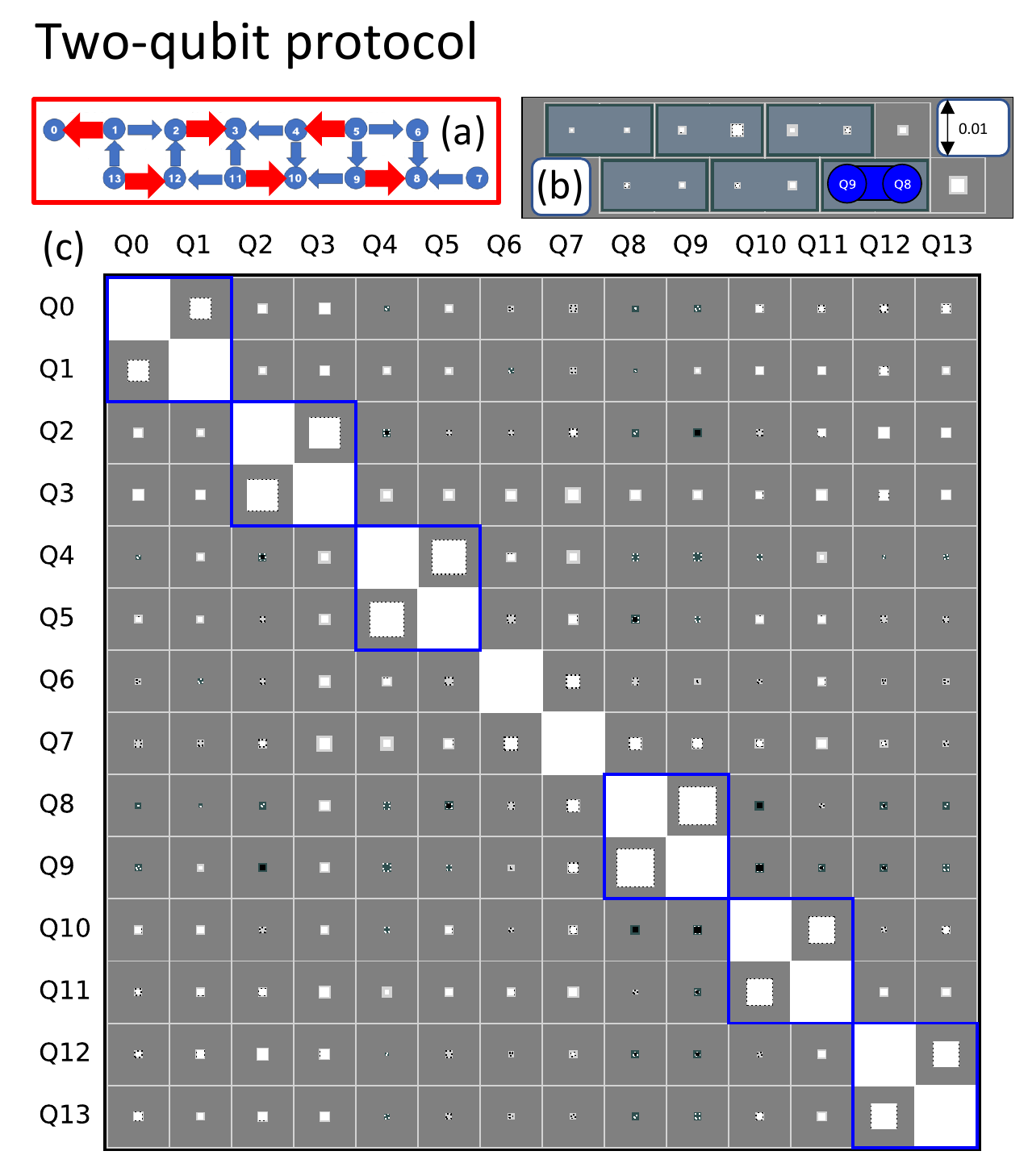}
\caption{\label{fig:doubleCorrelation}Results from the protocol run in two-qubit mode. (a) The two-qubit protocol partitions the qubits into disjoint sets of one or two qubits. 
Here we present one of the partitionings used in the experiments (see supplementary information for others).
(b) The chart shown is a Hinton diagram, laid out to reflect the physical layout of the device. 
The  mutual information between a qubit and the qubit pair Q8 and Q9 is shown as a white square, the area of which is proportional to the value. 
A full square represents a value of 0.01 Shannons (= bits $\times \ln 2$).
By treating the qubit pair as a single variable we remove the mutual information between the paired qubits, allowing us to see the mutual information captured between the qubit pair and the other qubits.
The qubit pairings are indicated with blue shading. 
As can be seen there is significant mutual information between qubit 3 and this particular qubit pair. 
This type of correlation is longer than nearest-neighbor and would be difficult to detect with previous protocols. 
(c) The correlation matrix computed from the reconstructed global probability distribution corresponding to the distinct qubit-pair groupings shown in (a). 
The blue boxes show the qubit couplings.
One would expect each qubit in a blue box to have the same correlations with all other qubits.
We observe that this is not the case, as, for example, qubit 4 is anti-correlated with qubit 2 but correlated with qubit 3.
We believe that this arises due to non-Markovian errors, but we leave a more detailed investigation for future research.
While there are a number of long-range interactions that are significant, it is easy to identify that the errors on qubit 11 positively correlate with the errors on each other qubit in a way that was not occurring with the single qubit protocol. The JSD between the probability distribution measured using the the two-qubit protocol and the local GRF model has increased by almost an order of magnitude to $0.216(1)$. This confirms that in two qubit mode the candidate GRF does not describe the system well, indicating the existence of many long range correlations. All error bars shown are 1$\sigma$ bounds, illustrated using the convention explained in \cref{fig:singleCorrelation} }
\end{figure}

To achieve an efficient protocol, scaling polynomially in $n$, we introduce the notion of a \emph{Gibbs random field} to describe the error rates $p(j)$.
A Gibbs random field (GRF) is a strictly positive probability distribution that obeys certain conditional independence properties known as Markov conditions (see Methods).
These conditions restrict the range of possible correlations enough to make the problem of noise characterization tractable, but they allow sufficient expressive power that a GRF can accurately model noise in real devices. The underlying (but testable) assumption is that realistic devices will only have correlations between a bounded number of qubits.

\begin{figure}[t!]
\includegraphics[width=0.99\columnwidth]{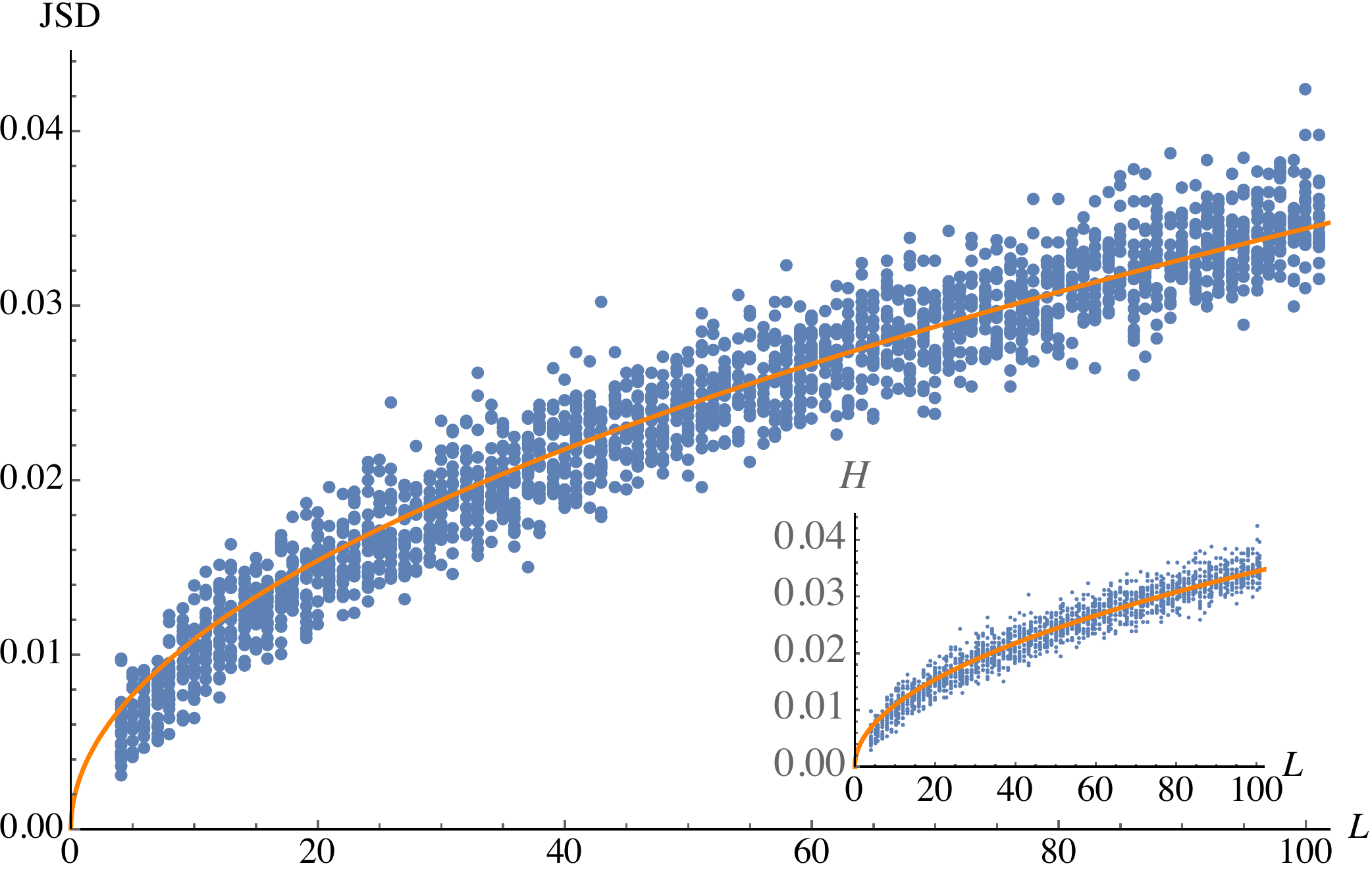}
\caption{\label{fig:JSDHellinger}
Illustration of the applicability  of the protocol to larger systems. The Jensen-Shannon distance (JSD) and Hellinger distance ($H$) (see Methods) for comparing the true ($p$) and estimated ($q$) probability distribution of Pauli errors on a chain of length $L$ for $L \le 100$, with $20$ random instances per value of $L$. 
The simulated noise is modeled by nearest-neighbor correlations in some true distribution $p$, and the estimate is reconstructed using the method in the text by estimating local marginal probability distributions to build a global estimate $q$ in a factorized form. 
Each local marginal was estimated using a constant number of samples, independent of $L$, so that the growth in the error with $L$ measures how global noise reconstruction accuracy degrades when the accuracy of the local estimates is fixed. 
The orange fit lines are the result of single-parameter least-squares fits to the model $a \sqrt{x}$.
The good qualitative agreement indicates that errors from local statistical fluctuations add up incoherently rather than constructively.}
\end{figure}

When quantum noise is approximated by a GRF, the parameters of the GRF can be learned efficiently, precisely, and accurately by only estimating the marginal distributions on the factors and their neighbors. 
When the factors have a small bounded size, then the protocol detailed in \autoref{fig:protocol} applied to the subsets of qubits for each factor performs this estimation with aplomb. 
The specific methodology for estimating the global probability distribution as a GRF from the estimated marginals is discussed in the Methods and the supplementary information. 
Note that it is not necessary to know the topology of the GRF factor graph in advance, rather it should be possible to learn the topology of a GRF that adequately describes the observed probabilities from the data itself using techniques from classical machine learning~\cite{Koller2009}. 
We discuss this briefly in the supplementary information but it remains an open problem, although we aim to show how to learn the topology in future work.

Our first experiments were run using the single-qubit protocol on the 14-qubit superconducting quantum architecture \textit{Melbourne}, made available by IBM through the IQX online quantum computing environment~\cite{Qiskit}. 
After completion of stage 4 of the protocol, we had reconstructed the entire averaged noise on the machine, returning all the averaged eigenvalues with multiplicative precision.

The real utility of the protocol then comes from its unique ability to reconstruct the SPAM-free qubit error rates. 
This allows us to utilize the standard tools for analyzing probability distributions to understand the noise correlations in the device. 
Indeed, any functional of the reconstructed probability distribution can be computed from the data, such as the mutual information between pairs of qubits, the covariance matrix of the errors, or the correlation matrix (as in \autoref{fig:singleCorrelation} or \autoref{fig:doubleCorrelation}).
In particular, the covariance and correlation matrices can be computed unconditionally in polynomial time irrespective of any efficient GRF description by using our protocol (see Supplementary Information). 
These tools provide invaluable diagnostic information about correlated errors that is difficult or impossible to obtain using prior art. 

Of critical importance in developing error-corrected quantum devices will be the identification and elimination of long-range qubit correlations. 
Our protocol allows such correlations to be identified and quantified. 
A GRF model that enforces short-range correlations can be constructed using the topology of the connections for the device in question. 
Reconstructing the observed probability distribution within the ansatz of the GRF allows two candidate distributions with different locality of their noise to be compared using metrics such as the Jensen-Shannon distance (JSD) between the distributions (see Methods). 
If calculated using $\log_2$, the JSD is bounded by $0$ and $1$.
If two candidate distributions $p(j)$ and $q(j)$ for describing the noise in the device have different factor topologies, but are close in JSD, then the model with more elaborate connections is likely overfitting to the data, and we can safely reject the additional correlations that the model represents as potential factors. 
In this way, multiple reconstructions can allow us to test for specific correlations in the device. 

\autoref{fig:singleCorrelation} illustrates the information that can be learned from the probability distribution, the construction of a short-range GRF model for the noise and the comparison of the GRF model to the global estimated distribution. 
In this case we calculated the JSD to be $0.042(8)$, demonstrating that when the device is operated using only single-qubit gates the long-range qubit correlations are small for these operations, although the non-zero JSD confirms that some long-range qubit correlations do exist. 
This is an important fact when considering such things as error correction and helps explain recent successful demonstrations of important elements of fault tolerance on the device~\cite{Harper2018}.

The two-qubit protocol, which activates resonators simultaneously between qubit pairs, unsurprisingly gives an entirely different characterization of the device. We can avoid the strong mutual information between qubits in each pair from swamping smaller correlations to other qubits by treating paired  qubits as a unified pair (i.e.~a single random variable). We then calculate metrics such as the mutual information between the qubit pair and other qubits (or pairs of qubits). 
This is illustrated in \autoref{fig:doubleCorrelation}.

While \textit{unknown} correlations would adversely affect the performance of error correction routines on the device, here we have fully calculated the averaged noise afflicting the system. 
This will allow tailored decoders to be constructed that can utilize the noise profile to increase the probability of successfully decoding error syndromes.
These correlations also provide valuable diagnostic information that can be used to learn the microscopic origin of the noise and potentially mitigate the source of the errors.

Finally, while the experiments here show the efficacy of this protocol on 14 qubits, our numerical simulations confirm the accuracy with which the protocol reconstructs the global estimates on much larger systems. 
In \autoref{fig:JSDHellinger} we present a numerical simulation demonstrating the accuracy of the protocol in reconstructing a GRF for a system with up to 100 qubits in a line by measuring few-body marginal distributions with a fixed number of samples.
The size of the simulated systems is far beyond what any other method is currently capable of characterizing.
The supplementary information contains more examples of simulations showing that the methodologies discussed here are indeed accurate and scalable.
Mathematical proofs providing recovery guarantees up to relative precision using this protocol are published in a separate paper~\cite{Flammia2019}. 
These results also prove that, subject to mild caveats, this protocol runs in polynomial time in the number of qubits. 

Our experiments give the first demonstration of a protocol that is practical, relevant, and immediately applicable to characterizing error rates and correlated errors in present-day devices with a large (>10) number of qubits. 
This protocol opens myriad opportunities for novel diagnostic tools and practical applications.
For example, structure learning  heuristics can be applied to try to learn the most accurate and efficient GRF noise model that describes the error rates.
In addition to the applications mentioned in the abstract, machine learning for fine-tuned error-correction decoders using the actual noise map of the device, quantum control for optimal gate synthesis, and noise-aware circuit compiling techniques are just some of the applications of this new method for characterizing quantum noise.

\textbf{Acknowledgements.} 
We thank S.~Bartlett, A.~Doherty, J.~Emerson, and T.~Monz for comments on an earlier draft.
This work was supported in part by the US Army Research Office grant numbers W911NF-14-1-0098 and W911NF-14-1-0103, the Australian Research Council Centre of Excellence for Engineered Quantum Systems (EQUS) CE170100009, the Government of Ontario, and the Government of Canada through the Canada First Research Excellence Fund (CFREF) and Transformative Quantum Technologies (TQT), the Natural Sciences and Engineering Research Council (NSERC), Industry Canada.

\textbf{Author contributions.} 
RH, STF, and JJW conceived the experiments, and STF and JJW conceived the original methodology. 
The implementation was carried out by RH. 
RH wrote the initial draft and all authors contributed to the revisions and editing of the manuscript. 

\textbf{Competing interests.}
JJW is the chief technology officer of the company Quantum Benchmark, Inc., and 
STF and RH were both consultants for it for part of the duration of this project.

\textbf{Additional information.} 
Supplementary information accompanies this paper on [weblink to be inserted by editor].

%

\newpage
\section{Methods}\label{sec:methods}

\textbf{Analysis of probability distributions.}
In the paper we use various metrics to analyse the probability distribution returned by the protocol.
Here we formally define the terms we use.
The \textit{relative entropy}, also known as the Kullback-Leibler (KL) divergence, between two probability distributions is one measure, and is defined as 

\begin{align}
    D(p\|q) = \sum\limits_{j}p(j)\ln \frac{p(j)}{q(j)} \,.
\end{align}

The \textit{mutual information} is a measure of the dependence between two random variables $X$ and $Y$, and it quantifies the amount of information obtained regarding one variable through observing the other.
It is defined as:

\begin{align}
    I(X,Y) &= D\bigl(p(x,y)\|p(x)p(y)\bigr)\\
    &=\sum\limits_{x,y}p(x,y)\ln\frac{p(x,y)}{p(x)p(y)}\,,
\end{align}

where $D\bigl(p(x,y)\|p(x)p(y)\bigr)$ is the relative entropy between the joint probability distribution $p(x,y)$ and the marginal distribution on each of the random variables.

For each run of the protocol, we have a partitioning of the set of 14 qubits into disjoint sets $s$ that are acted upon by independent twirls.
For each set $s$ we define a random variable $Q_s$ that takes on the value 0 if no error acts on the set $s$ and 1 otherwise.
For the single-qubit protocol, we have 14 random variables $Q_0,\ldots,Q_{13}$.
We then calculate the mutual information between each pair of random variables.
For the two-qubit protocol, the set $s$ comprises one or two qubits, depending on whether two-qubit gates were used on that pair or not.

\textit{Conditional mutual information} represents the expected value of the mutual information of two random variables conditioned on the value of the third.
In the present case we have:

\begin{align}
    &I(X,Y|Z) \nonumber\\
    &= \sum\limits_{z\in Z}p(z)\sum\limits_{y\in Y}\sum\limits_{x\in X}p(x,y|z)\log\left(\frac{p(x,y|z)}{p(x|z)p(y|z)}\right)\\
        &= \sum\limits_{z\in Z,y\in Y, x\in X}p(x,y,z)\log\left(\frac{p(z)p(x,y,z)}{p(x,z)p(y,z)}\right)\,.
\end{align}

One can use the probability distribution to compute the \textit{covariance} matrix between the observed error random variables.
In the experiment presented in this paper, where we have observed errors, we can treat the qubits as 0/1 random variables representing no error/error.
Then, if $x$ is a column vector representing an error pattern, we compute the covariance matrix $\mathbf{\Sigma}$ as

\begin{align}
   \mathbf{\Sigma}= \mathbb{E}\left[(x-\mu)(x-\mu)^T\right]\,,
\end{align}
where $\mu=\mathbb{E}\left[x\right]$ and $\mathbb{E}$ denotes the expected value over the probability distribution.

Given the covariance it is simple to calculate the Pearson product-moment correlation coefficient matrix ($Q$), obtained by dividing the covariance of the two variables by their standard deviation.
This is the \textit{correlation matrix} plotted in the paper.
In this case let $V=\mathrm{diag}(\mathbf{\Sigma})$. 
Then we have:

\begin{align}
    Q = V^{-\frac{1}{2}}\,\mathbf{\Sigma}\, V^{-\frac{1}{2}}\,.
\end{align}

In the case where one averages over the Pauli group (instead of the Clifford group) to characterise the Pauli noise of the device (rather than the basis-averaged Pauli noise of the device presented here in our experiments), then the relevant qubit random variable can be characterized by the $2n\times 2n$ Pauli covariance matrix, where each row and column of the matrix is labeled by the $2n$-bit string required to label each of the $\sigma_x$ and $\sigma_z$ components of a single-qubit Pauli on $n$ qubits.

One of the problems with the relative entropy $D(p\|q)$ is that it is defined only if $\forall j, q(j)=0\implies p(j)=0$ and this might not be the case in practice, especially when numerical precision issues are taken into account.
For this reason we use the Jensen-Shannon distance ($\mathrm{JSD}$) which is defined as:

\begin{align}
    \mathrm{JSD}(p,q)&=\biggl(\frac{1}{2}D(p\|m)+\frac{1}{2}D(q\|m)\biggr)^{1/2}\,,
\end{align}
where $m(j) =\bigl(p(j)+q(j)\bigr)/2$.
This is a smoothed, symmetric measure which always has a finite value and has the mathematical properties of a metric.

In \autoref{fig:JSDHellinger} we use the \emph{Hellinger distance}. 
The Hellinger distance forms a bounded metric on the space of probability distributions over a given probability space and, for discrete distributions, is defined as:

\begin{align}
    H(p,q) = \biggl(1-\sum_j \sqrt{p(j) q(j)}\biggr)^{1/2}\,.
\end{align}

This distance is efficient to compute and is related to the more commonly used but difficult to compute notion of \emph{statistical distance} (or total variational distance) $\delta(p,q)$ as:

\begin{align}
H^2(p,q)\le \delta(p,q)\le \sqrt{2}H(p,q)\,,
\end{align}
where 
\begin{align}
    \delta(p,q) = \frac{1}{2}\sum_j |p(j)-q(j)|\,.
\end{align}

\textbf{Gibbs Random Fields.}

Associated to any Gibbs random field (GRF) is a factorization of the error rates into a product of factors, where each factor is a positive function depending only on a subset of the qubits. 
The \emph{factor graph} of this factorization, depicted in \autoref{fig:singleCorrelation}, has two types of nodes: one set for the random variables associated to each single-qubit error, and one set for each factor in the factorization of $p$, with factor nodes connected to exactly the qubit nodes they depend on. 
We say that two single-qubit nodes are adjacent if they connect to some common factor. 
The Markov conditions then say that the correlations in a GRF are of a specific form: errors on any subset of non-adjacent qubits are conditionally independent given the errors on their neighbors.

Accordingly, any GRF can be depicted as an undirected graphical model of a set of random variables having the property (known as the local Markov condition), that any subset of variables is conditionally independent of all other variables given its neighbors.

Here the notion of adjacent is computed with respect to the factor graph of the model. 
That is, for a given subset $A$ of random variables and a subset $B$ containing the boundary $\partial A$ of $A$, the following holds $\text{Pr}(x_A|x_B) = \text{Pr}(x_A|x_{\partial A})$, where $x_A$ is the random variable describing the errors on the qubits in $A$. 

For the current experiment we can use the topology of the device to define a Markov network, where the connections in the graph are identical to the resonators between the qubits.
The graph appears in \autoref{fig:singleCorrelation}.
If we wish to enforce short-range correlations then we have: 

\begin{align}
    &p(0|1,13,2,12,3,11,4,10,5,9,6,8,7)=p(0|1,13)\nonumber\\
    &p(1,13|0,2,12,3,11,4,10,5,9,6,8,7)=p(1,13|0,2,12)\nonumber\\
    &p(2,12|0,1,13,3,11,4,10,5,9,6,8,7)=p(2,12|1,13,3,11)\nonumber\\
    &p(3,11|0,1,13,2,12,4,10,5,9,6,8,7)=p(3,11|2,12,4,10)\nonumber\\
    &\dots\,,\label{eq:markovblanket}
\end{align}
where we have have used numbers to represent random variables, e.g.~$0$ for $x_0$, to de-clutter the notation.

We can then use the chain rule to write the joint probability distribution as follows:

\begin{align*}
    p(0,1,13,2,12,3,11,4,10,5,9,6,8,7) =\ \ \ \  \\
    p(0|1,13,2,12,3,11,4,10,5,9,6,8,7)\times\\
    p(1,13|2,12,3,11,4,10,5,9,6,8,7)\times\\
    p(2,12|3,11,4,10,5,9,6,8,7)\times\\\dots\times\\ p(6,8|7)\times p(7)\,
\end{align*}
then using \autoref{eq:markovblanket}, this simplifies to:
\begin{align*}
    p(0|1,13)p(1,13|2,12)p(2,12|3,11)\dots p(6,8|7)p(7)\,
\end{align*}
which we can calculate as
\begin{align}
    &\frac{p(0,1,13)}{p(1,13)}\times\frac{p(1,13,2,12)}{p(2,12)}\times\frac{p(2,12,3,11)}{p(3,11)}\nonumber\\
    &\dots\times\frac{p(6,8,7)}{p(7)}\times p(7).\label{eq:induced}
\end{align}

Finally we can use the probability distribution defined in \autoref{eq:induced} to reconstruct the \textit{Gibbs random field}, that is, the probability distribution induced by the condition imposed by our chosen Markov conditions.
A comparison between the observed probability distribution and the induced one by, say, the Jensen-Shannon distance gives us a metric to observe how closely our device corresponds to a device that only has short range correlations.

\textbf{IBM Quantum Experience.}
The experiments reported here were conducted on the IBM Quantum Experience \textit{Melbourne} device.
Jobs were submitted via Qiskit~\cite{Qiskit} in two separate runs.
The single qubit experiment consisted of 1,000 different submissions.
Each submission contained 11 single-qubit Clifford twirls on each of the 14 qubits for gate lengths $1, 5, 10, 15, 20, 30, 45, 60, 75, 90, 105$, with each gate length sequence requesting 1024 shots.
This meant in total $1024000\times 11$ measurements were made.
For step 4 of the protocol (see \autoref{fig:protocol}) simple least squares fitting was used, although for any sequence length where the value was less then $(p_1+1/16)/4$ that data and the data from longer sequences were discarded for the purposes of the fit.

With the two-qubit protocol the topology of the \textit{Melbourne} device does not allow seven qubit pairs to be operated simultaneously.
The qubits were grouped into six distinct qubit pairs, the remaining two qubits being operated in single-qubit-protocol mode.
In order to determine which qubit pairs to activate, attention was paid to the reported fidelities of the CNOT gates.
For instance on 18 February 2019, the two-qubit gate between qubits 13 and 1 had a reported fidelity of $84.1\%$ (other gates could be as high as~$97\%$).
The groupings shown in \autoref{fig:doubleCorrelation} were chosen to attempt to ensure wide coverage of the various two-qubit gates available while avoiding any gates with a reported fidelity close to or below $90\%$. 
Only one of the three configuration settings is reported in the paper, the others appear in the supplementary information. 
Because of the possibility that correlations within the device might change with re-calibration all the different runs shown in \autoref{fig:doubleCorrelation} were interleaved.
In total there were $1924$ different submissions (of 11 different sequence lengths) for each of the three different configurations.
One re-calibration cycle did occur during the gathering of the data, although the fidelities of the two-qubit gates did not appear to change significantly as a result.
The data does, however, represent an average of the noise that occurred in the machine during that time period.
Given the reduced fidelity of the runs (since two-qubit gates have an infidelity of an order of magnitude greater than the single-qubit gates) sequence lengths were reduced to $0\dots10$.
The 0 sequence, representing one single qubit Clifford (i.e.~no two-qubit gate), was added so as to allow a more accurate determination of the $A$ constant in the fit.
As with the single-qubit runs, data with a value less than $(p_0+1/16)/4$ were discarded for the purposes of the fit, although a minimum of three data points were retained.

Finally in all cases an $X$ gate was randomly compiled into the qubits of each sequence submitted, with the probability distribution interpreted accordingly so as to eliminate any bias in the SPAM~\cite{Harper2019}.

\textbf{Error Bars.}
All error bars shown here were calculated using non-parametric bootstrap methods.
For each sequence length of each run the observed probability distributions of the measurement counts (Step 2 in the protocol shown in \autoref{fig:protocol}) were re-sampled (with replacement) for the same number of measurements used to originally gather the data.
This was repeated 1,000 times.
Each of these 1,000 sets of re-sampled distributions were then analyzed in a manner identical to the original (steps 3, 4 and 5 of the protocol), to provide 1,000 bootstrapped samples of the SPAM free probability vector (the \textit{bootstrapped probabilities}).
From this the appropriate confidence intervals to provide error bars can be constructed.
With the mutual information estimates the mutual information between the qubits in question can be calculated 1,000 times, ordered, and by extracting the values at the appropriate location of the ordered values, the confidence intervals are obtained (so for $1\sigma$ confidence intervals the 159th and 841st values are used).
With the error bars on the correlation matrices the following conservative approach was adopted.
Using the bootstrapped probabilities, 1,000 correlation matrices were constructed.
Since there is no clear way to order such matrices, each individual cell on the matrices was treated separately, with the possible values for that cell location being ordered and the appropriate high/low values being extracted as before.
While a matrix constructed from all the low (or high) values would not in itself be a valid correlation matrix, it is believed the error bars still, conservatively, convey the confidence intervals for each of the individual values in the matrix.
Finally with the calculation of the Jensen-Shannon distance two different re-sampling techniques were utilized.
In the first the Gibbs random field reconstructed probability distribution for the originally observed distribution was compared to each of the bootstrapped probability distributions and in the second a Gibbs random field distribution was constructed from each bootstrapped probability distribution and compared to the full bootstrapped distribution it was constructed from.
In all cases the error ranges were broadly similar and in the paper the uncertainties quoted were taken from largest error ranges from either of the methodologies.

\textbf{Data Availability}

Source data are available for this paper at
\begin{itemize}
    \item \url{https://github.com/rharper2/EfficientLearningDataSet} and
    \item \url{https://doi.org/10.5061/dryad.q83bk3jf0}. 
\end{itemize}

All other data that support the plots within this paper and other findings of this study are available from the corresponding author upon reasonable request.

\textbf{Code Availability}

Details on code that was used to analyse the data is available from the corresponding author upon reasonable request. Jupyter notebooks recreating the graphs can be found at \url{https://github.com/rharper2/Juqst.jl} (docs/examples/quantumNoise) and code to create similar circuits to those used and to submit them to the IBM device at \url{https://github.com/rharper2/query_ibmq}

\clearpage

\part*{Supplemental Information}
\begin{figure*}[ht!]
\begin{flushleft}
(a)
\end{flushleft}

\includegraphics[width=0.9\textwidth]{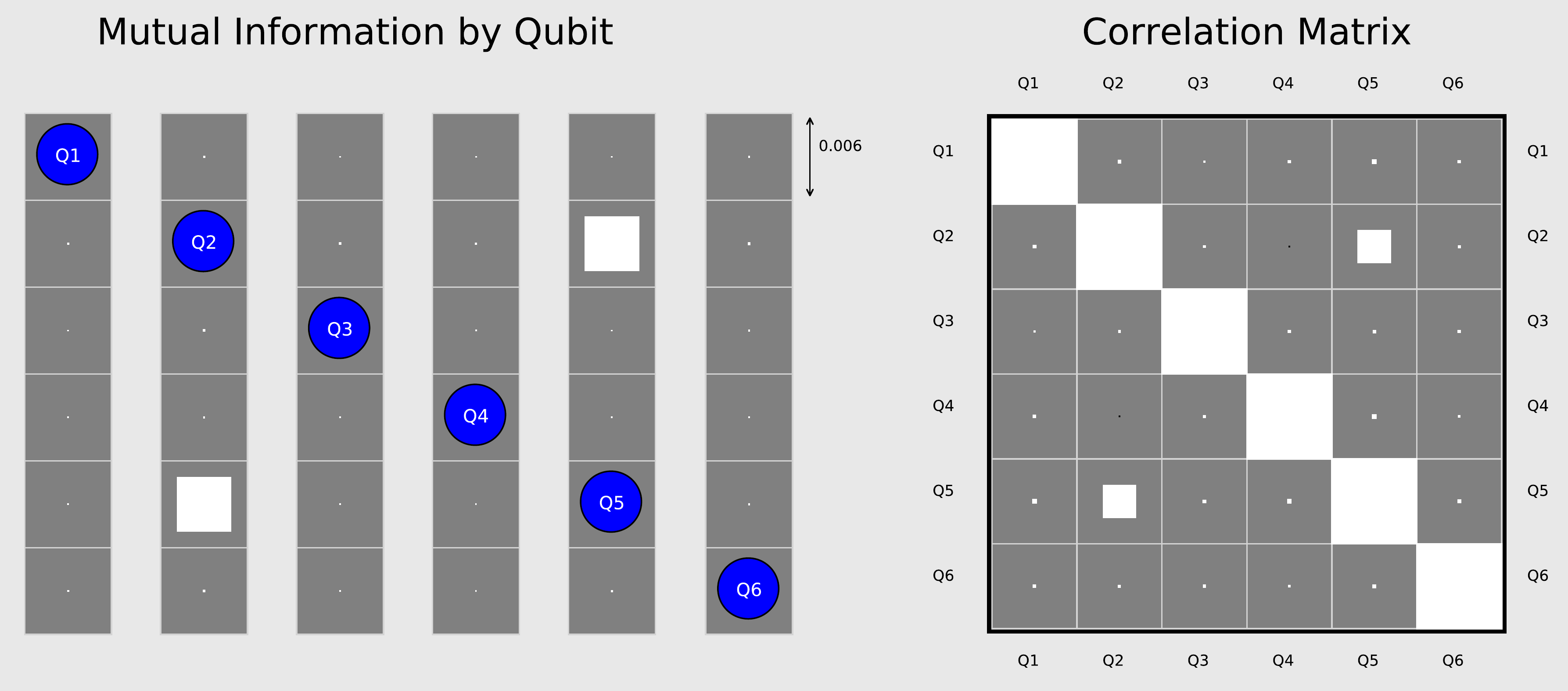}

\begin{flushleft}
(b)

\end{flushleft}

\includegraphics[width=0.9\textwidth]{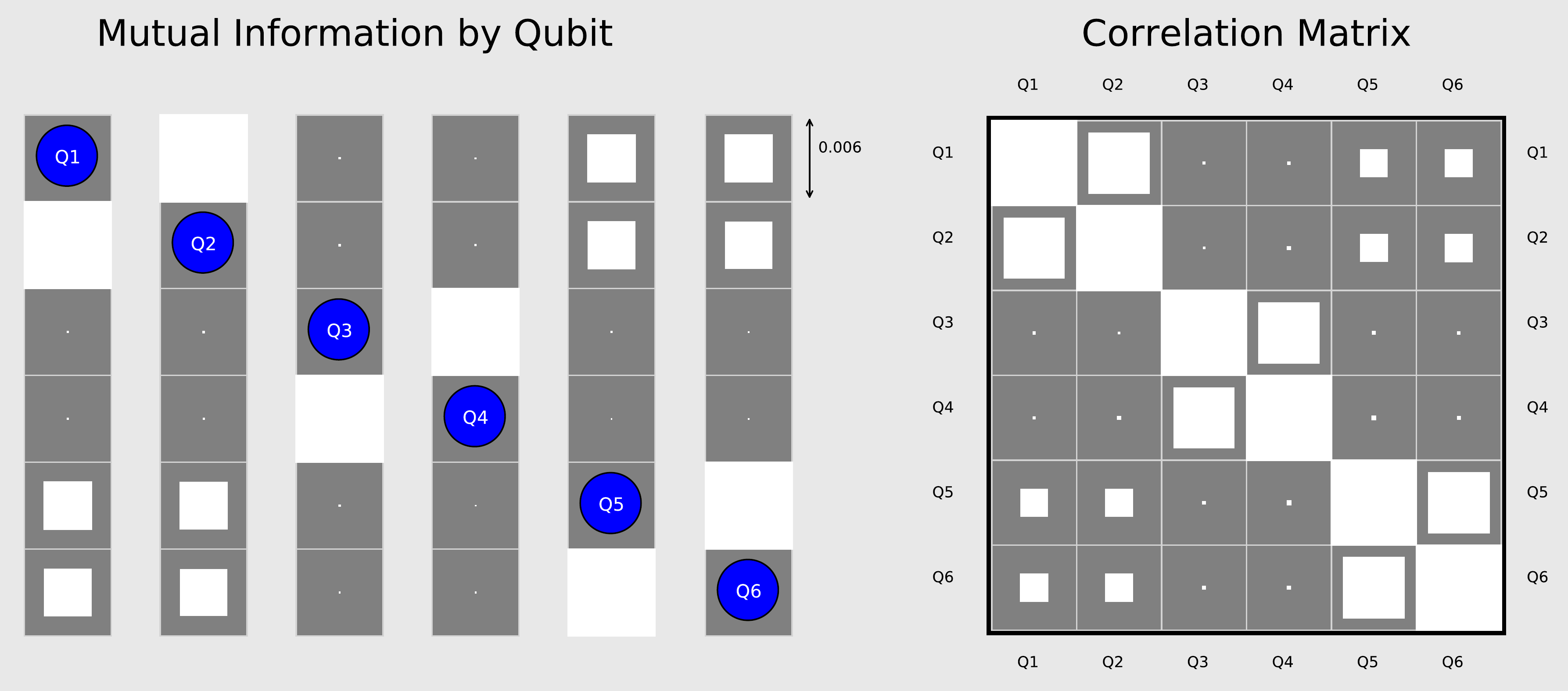}

\caption{Numerical simulation of a noisy 6 qubit system. 
Left hand plots show the mutual information between each identified qubit and the others via a Hinton diagram. 
Right hand plots show a correlation matrix identifying the correlations between each of the qubits. 
Relative size of white/black represents the strength of the mutual information/correlation, with the color white (black) being positive (negative). 
(a) The plots here arise from a simulation where the noise matrix is tensored single qubit noise with an additional two-qubit interaction between qubits 2 and 5. 
The protocol is operated in single qubit mode (see \autoref{sec:singleQubitProtocol}). 
As can be seen the protocol allows the identification of the correlation. 
(b) The system here has single-qubit noise applied to each qubit. 
The protocol is operated in `two-qubit' mode (\autoref{sec:twoQubitProtocol}) with the qubits being paired as (1,2), (3,4) and (5,6). 
Each time a two-qubit gate is activated additional noise is applied to the qubits involved representing noisy two-qubit gates. 
Finally when the two-qubit gate between qubits 1 and 2 is activated further crosstalk noise is applied as between qubits 2 and 5 (representing crosstalk between the resonator between qubits 1 and 2 and the operation of qubit 5). 
Since errors are transmitted through the two-qubit gates of paired qubits there is significant mutual information and correlations between such qubits. 
The charts show the additional crosstalk (2 to 5) as mutual information (or correlations) between qubits 1,2 and 5,6. 
(The two-qubit operation spreads this error from 5 to 6). 
See \autoref{sec:fidelityExtraction} for fidelity extraction.}\label{fig:simulation}
\end{figure*}

\setcounter{section}{0}

\section{Preliminaries and Notation}
\subsection{Channels}
We analyze noise as quantum channels, that is, as completely positive trace preserving (\textbf{CPTP}) linear maps $\mathcal{L}:\mathbb{C}^{d\times d}\rightarrow \mathbb{C}^{d\times d}$, mapping operators to operators.
For $n$ qubits, we have $d=2^n$.

It is convenient to work in the superoperator representation of quantum channels. 
The superoperator representation is conveniently expressed in a orthonormal operator basis (according to the Hilbert-Schmidt inner product $\mathrm{Tr} A^\dagger b$). 
Here we will use the Pauli basis representation of a channel $\mathcal{E}$ in which the basis is the set of suitably normalized Pauli operators.
For a single qubit, we choose the basis to be $\mathbb{A} = \{I, X, Y, Z\}/\sqrt{2}$.
For systems with $n$ qubits, the orthonormal basis is formed from the natural tensor products of these four Pauli operators, with $A_0$ being the appropriately normalized $\mathbbm{1}^{\otimes n}$.

The superoperator representation of a quantum channel acts naturally via matrix multiplication on the vectorization of a density operator $\rho$, which we denote by $|\rho)$.
The notation $|\rho)$ simply indicates that $\rho$ is now being treated as a vector whose components can be expanded in the above-mentioned orthornormal operator basis (in this case the Pauli basis). 
Concretely, we can expand the density operator as $\rho = \sum_k \langle A_k,\rho\rangle A_k$, where $\langle A_k,\rho\rangle$ represents the Hilbert-Schmidt inner product $\mathrm{Tr}(A_k^\dagger \rho)$ between the basis element $A_k$ and the density matrix $\rho$. 
Once we have done that we can identify $\rho$ with the a column vector $|\rho)\in\mathbb{C}^{d^2}$ where the k\textsuperscript{th} entry is the relevant Hilbert-Schmidt inner product specified above. 
The superoperator representation of a channel $\mathcal{E}$ is then the unique matrix $\mathcal{E}\in\mathbb{C}^{d^2\times d^2}$ such that $\mathcal{E}|\rho)=|\mathcal{E}\left(\rho\right))$.

When we have fixed $A_0$ to be the rescaled identity operator, any CP channel $\mathcal{E}$ can be written in block form as 

\begin{align}\label{eq:block_terms}
\boldsymbol{\mathcal{E}} = 
\begin{pmatrix}
S(\mathcal{E})  & \boldsymbol{\mathcal{E}}_{\mathrm{sdl}} \\ 
\boldsymbol{\mathcal{E}}_{\mathrm{n}} & \boldsymbol{\mathcal{E}}_{\mathrm{u}}
\end{pmatrix},
\end{align}
where we refer to $\boldsymbol{\mathcal{E}}_{\mathrm{sdl}}$, $\boldsymbol{\mathcal{E}}_{\mathrm{n}}$ and 
$\boldsymbol{\mathcal{E}}_{\mathrm{u}}$ as the \textit{state-dependent leakage}, \textit{nonunital} 
and \textit{unital} blocks respectively (see \cite{Wallman2015a} for more details about this decomposition). 
If the channel is trace preserving then $\boldsymbol{\mathcal{E}}_{\mathrm{sdl}}=0$. 
The unital block ($\boldsymbol{\mathcal{E}}_{\mathrm{u}}$), and more precisely the diagonal of the unital block, contains all the information necessary to extract the Pauli noise afflicting the system, including for instance the fidelity of the channel and/or the fidelity of any subspace of the system.

From this representation, the \emph{Pauli projection} is the projection of the channel to just its diagonal entries (in the Pauli basis). 
In particular, this projection can be effected by \emph{twirling} the channel with respect to the Pauli group, i.e., by averaging the noise channel as $\mathcal{E^P}(\cdot) := 4^{-n}\sum_j P_j\mathcal{E}(P_j \cdot P_j)P_j$. 
The diagonal elements of $\boldsymbol{\mathcal{E}}$ or of the Pauli projection $\boldsymbol{\mathcal{E}}^\mathcal{P}$ are exactly the Pauli eigenvalues, $\lambda(j) = 2^{-n}\mathrm{Tr}\bigl(P_j \mathcal{E}(P_j)\bigr)$.

When the noise channel is twirled by the Clifford group, the resultant noise channel will also have $\boldsymbol{\mathcal{E}}$ diagonalized in the Pauli basis, with each element being the average of the diagonal elements of the original $\boldsymbol{\mathcal{E}}_{\mathrm{u}}$. 
By way of illustration in the single-qubit case, a completely positive trace preserving noise channel will be of the form:

\begin{align}\label{eq:general_noise}
\boldsymbol{\mathcal{E}} = \begin{pmatrix}
1 & 0 & 0 & 0\\ 
\alpha_1 &\lambda_x &\delta_1 &\delta_2\\ 
\alpha_2 &\delta_3 &\lambda_y &\delta_4 \\ 
\alpha_3  & \delta_5 &\delta_6 & \lambda_z
\end{pmatrix}
\end{align}
where all the elements are real. 
The matrix elements themselves obey certain constraints on account of the requirement for complete positivity; see Ref.~\cite{King2001} for an explicit description of these constraints. 
In particular we note that for purely decoherent noise, $\delta_{i} =0$ for all $i=\{1..6\}$, whereas for purely coherent noise acting on $n$ qubits we have $\mathrm{Tr} \boldsymbol{\mathcal{E}}_{\mathrm{u}}^\dagger \boldsymbol{\mathcal{E}}_{\mathrm{u}} = 2^{2n} - 1$.

After averaging over the Clifford group, the unital block  $\boldsymbol{\mathcal{E}}_{\mathrm{u}}$ of the twirled error channel looks like

\begin{align}\label{eq:single_unital}
\begin{pmatrix}
\lambda_\Sigma &0&0\\
0&\lambda_\Sigma  &0\\
0&0& \lambda_\Sigma 
\end{pmatrix}\,,
\end{align}
where $\lambda_\Sigma = (\lambda_x + \lambda_y + \lambda_z)/3$.
If the average were over the Pauli group, the unital block would be:
\begin{align}
\begin{pmatrix}
\lambda_x&0&0\\
0&\lambda_y&0\\
0&0&\lambda_z
\end{pmatrix}\,.
\end{align} 

The standard randomized benchmarking protocol twirls the noise through the Clifford group, meaning that the depolarizing factor measured by it (the decay rate) is simply $\lambda_\Sigma$. 

More generally, RB over the full Clifford group will learn the decay parameter $\bar{f}:=\frac{1}{d^2-1} \sum_{j \not= \mathbbm{1}} \lambda_j$. 
The number typically reported by experiments, however, is the closely related notion of \emph{average gate fidelity} $\mathcal{F}$, which is a shifted and rescaled version of this given by $\mathcal{F} = \frac{(d-1) \bar{\lambda}+1}{d}$.
Some experiments report instead the average gate \emph{in}fidelity or \emph{average error rate} $r$, which is just $r = 1-\mathcal{F} = \frac{d-1}{d}(1-\bar{\lambda})$. 

When we have a Pauli-projected noise matrix we will refer to each of the diagonal entries in the matrix as a \textbf{decay parameter}, which we will typically denote as $\lambda$. 
For a Pauli channel, these diagonal elements are the channel eigenvalues, so we will refer to them as the Pauli eigenvalues. 
Unless otherwise clear by the context, $\lambda_x, \lambda_y, \lambda_z$, will refer to the $\lambda$ parameters associated with the specific sub-scripted Pauli channel, whereas $\lambda_\Sigma$ will typically refer to an $\lambda$ parameter associated with an averaged noise channel, such as in \cref{eq:single_unital}.

\subsection{Measurement}
\label{sec:measurement}
In the superoperator representation, a projective measurement is represented by a row vector for the projection operator expanded in the Pauli basis. 
So, for instance, in a two qubit system if we were to prepare and measure in the computational basis the entries would relate to the Pauli operators, $II$, $IZ$, $ZI$, and $ZZ$. 
Assuming each qubit is measured separately we can use $\uparrow$ to represent an `up' and $\downarrow$ a `down'. 
We then have four different possible measurement outcomes $\uparrow\uparrow\,, \uparrow\downarrow\,, \downarrow\uparrow\text{ and }\downarrow\downarrow$. 
The following relationship holds:

\begin{align}
&\uparrow\uparrow=\left(\begin{array}{c}II:+\\IZ:+\\ZI:+\\ZZ:+\end{array}\right)\,, \ \ 
\uparrow\downarrow=\left(\begin{array}{c}II:+\\IZ:-\\ZI:+\\ZZ:-\end{array}\right)\,,\nonumber\\
&\downarrow\uparrow=\left(\begin{array}{c}II:+\\IZ:+\\ZI:-\\ZZ:-\end{array}\right)\,, \ \ 
\downarrow\downarrow=\left(\begin{array}{c}II:+\\IZ:-\\ZI:-\\ZZ:+\end{array}\right)\,.
\end{align}

By inspection we can see that the (properly normalized) sum of the above gives the identity (i.e. we are asserting that the qubits will always be in one of these four states). 
More importantly we can rearrange the above to note that:

\begin{align}
    II &=\, \uparrow\uparrow + \downarrow\uparrow + \uparrow\downarrow + \downarrow\downarrow\nonumber\\
    IZ &=\, \uparrow\uparrow + \downarrow\uparrow - \uparrow\downarrow - \downarrow\downarrow\nonumber\\
    ZI &=\, \uparrow\uparrow - \downarrow\uparrow + \uparrow\downarrow - \downarrow\downarrow\nonumber\\
    ZZ &=\, \uparrow\uparrow - \downarrow\uparrow - \uparrow\downarrow + \downarrow\downarrow\,.\label{eq:hadamardTransform}
\end{align}
In other words we can reconstruct the relevant entries in a vectorized state if we know the percentage chance of each of the four different measurement outcomes. 
Clearly such outcomes will only be known approximately and we discuss later how to deal with this and how to relate the values to the relevant noise channel acting on a noisy state.

As shown in Ref.~\cite{Flammia2019} the relationship detailed in \cref{eq:hadamardTransform} is an inverse Walsh-Hadamard transform from the observed probability distribution to the relevant values of the noise matrix generating such a distribution. 
This relationship generalizes to $n$-qubits. 
Accordingly, given the probability distribution of observed outcomes over $n$-qubits one can perform a Walsh-Hadamard transform to determine the relevant entries in the state that produced them.

\subsection{Clifford subsystem twirling}
As noted in \cite{Gross2007a}, any unitary 2-design, which in particular includes the Clifford group, has exactly 2 irreducible representations  (\textbf{irreps}) in its matrix representation in the superoperator representation. 

Given a representation ($\phi,V$) defined by a homomorphism $\phi$ and a vector space $V$ for a group $\mathfrak{G}$ and a matrix $A \in \text{GL}(V)$, we define an action (the \textit{twirl}) 
\begin{equation}
A^g = \phi(g)\,A\,\phi(g^{-1})\,,
\end{equation} 
where $g^{-1}$ is the inverse of $g$. We can do this for each element of $\mathfrak{G}$, which gives us the uniform average of this action defined as:
\begin{equation}
A^G=\frac{1}{|\mathfrak{G}|}\sum_{g \in \mathfrak{G}}A^g,
\end{equation}
where $|\mathfrak{G}|$ is the number of elements in the group. 
The important thing to notice here is that $A^G$ commutes with the action of $\mathfrak{G}$ for any representation ($\phi,V$). 
Schur's Lemma for algebraically closed fields provides that if ($\phi,V$) is a finite-dimensional irrep and $\lambda$ is an intertwining operator, then $\lambda$ = $\eta \mathbbm{1}$ for some scalar $\eta$. 
In this case the action $A^G$ commutes with all representations and therefore is an intertwining operator.
In the case that the decomposition into irreps of a given representation is multiplicity-free, then the twirling operation makes any operator especially simple.
When a `twirl' is performed by averaging the noise matrix by the action of the group with such a multiplicity-free representation, the noise matrix in the superoperator basis is reduced to a diagonal matrix with the number of distinct entries equal to the number of irreducible representations of the twirling group.

For instance with a single qubit the noise matrix will have one parameter representing the trivial irrep (which will always be 1 for a quantum channel), with the second irreducible representation diagonalizing the matrix as shown in \cref{eq:single_unital}.  
In general where a subsystem is twirled by the full Clifford group (or, indeed, any unitary 2-design) corresponding to that subsystem, there are two irreducible sub-spaces corresponding to the two stabilized projectors, $\Pi_o$ (the trivial/identity operator) and a projector $\Pi_f$, being a projector onto the remainder of the Pauli group acting on those qubits. 
As shown in~\cite{Gambetta2012}, when we twirl over two distinct groups of subsystem Clifford operators (i.e. a $\mathcal{C}^{\otimes 2}$ twirl) we obtain four distinct irreducible sub-spaces $\Pi_0 = \mathbbm{1}\otimes\mathbbm{1}, \Pi_2 = \mathbbm{1} \otimes \Sigma, \Pi_1 = \Sigma \otimes \mathbbm{1} \text{ and } \Pi_{12} = \Sigma \otimes \Sigma$, where $\Sigma$ is the vector of Pauli operators for each subsystem. 
The size of each of the irreducible representations can be calculated from the character $\chi$ of the underlying irreducible representations.
In the case of a single qubit Clifford twirl this is $\chi_\mathbbm{1}=1,\chi_\Sigma=3$ and so for a $\mathcal{C}^{\otimes 2}$ we have:

\begin{align}
&\chi_{\mathbbm{1}\otimes\mathbbm{1}} = 1\times1 = 1\\
&\chi_{\Sigma\otimes\mathbbm{1}}=3\times1=3\\
&\chi_{\mathbbm{1}\otimes\Sigma}=1\times3=3\\
&\chi_{\Sigma\otimes\Sigma}=3\times3=9\,.
\end{align} 

This generalizes in the obvious way to twirls over $n$ qubits, where if there is a single Clifford twirl over each qubit, there will be $2^n$ irreducible representations with multiplicities given by simple multiplications of the characters.

Importantly in an $n$-qubit system where each qubit is independently twirled over the Clifford group, each irrep is of the form $\{\mathbbm{1},\Sigma\}^{\otimes n}$ and therefore each $\{I,Z\}^{\otimes n}$ Pauli belongs to one (and exactly one) of the irreps and the associated decay parameter (in the case of a twirled noise channel). 
To keep the notation as simple as possible where we refer to the decay parameter of these averaged channels with a $\Sigma$ in the appropriate place, omitting the $\otimes$ notation. 
So for a two qubit system, twirled with two independent Clifford twirls, the following holds:

\begin{align}
    \lambda_{II} = &\lambda_{II}\\
    \lambda_{\Sigma I} = &\frac{1}{3}(\lambda_{XI} + \lambda_{YI}+\lambda_{ZI})\\
    \lambda_{I\Sigma} = &\frac{1}{3}(\lambda_{IX} + \lambda_{IY}+\lambda_{IZ})\\
    \lambda_{\Sigma\Sigma} = &\frac{1}{9}(\lambda_{XX} + \lambda_{XY}+\lambda_{XZ}\nonumber\\
    &+\lambda_{YX} + \lambda_{YY}+\lambda_{YZ}\nonumber\\
    &+\lambda_{ZX} + \lambda_{ZY}+\lambda_{ZZ})
\end{align}

To summarize, if we have an $n$-qubit system and we perform a single-qubit Clifford twirl on each of the qubits, we diagonalize the noise matrix ($4^n\times4^n$ superoperator) representing the average difference between the noisy gates and their ideal counterparts \cite{Wallman2018} into a matrix that has $2^n$ different diagonal entries.
Furthermore, each entry is associated with a value that we can determine from the probabilities of measuring each qubit in the computational basis (see \cref{sec:measurement}).

\section{Learning the averaged noise}

\subsection{Randomized Benchmarking}
Here we recap the relevant parts of the randomized benchmarking protocol.

The \textit{raison d'\^etre} of randomized benchmarking is to allow small error rates to be measured in a way that is robust to state preparation and measurement (\textbf{SPAM}) errors in the device being characterized. 
It does this by using a Clifford twirl (discussed earlier) to transform the noise into a depolarizing channel with the same fidelity as the noise being twirled. 
At the end of each sequence a final inverting gate is applied (which would in an ideal system return the state of the system to its initial state).
One way of looking at this it that it allows a distribution relating to the probability of the state returning to the intended state to be sampled and the underlying percentages to be estimated in a SPAM-free manner. 
Given this, the typical randomized benchmarking protocol becomes:

\begin{enumerate}
\item Choose a positive integer $m$.

\item Choose a random sequence of gates $s$ from a set $\mathbb{S}_m$, typically of Clifford gates.
Note that in modern versions of this protocol these gates are chosen randomly to either leave the state invariant or to map it to an orthogonal state in order to eliminate a nuisance model parameter (see e.g.~Ref.~\cite{Harper2019}).

\item Obtain an estimate $\hat{q}(m,s)$ of the expectation value $q(m,s)$ of an observable $E$ after preparing a state $\rho$ and applying the gates in $s$.

\item Repeat steps 2--3 to obtain an estimate $\hat{q}(m)$ of $\bar{q}(m)$.
\item Repeat steps 1--4 and fit to the model

\begin{align}\label{eq:decay_model}
\bar{q}(m) = A c^m + B
\end{align}

where $c$ is related to some parameter of interest (e.g., the average gate fidelity to the identity) and $A$ is a SPAM-dependent constants and $B$ is a constant that varies with the number of qubits.
\end{enumerate}

This protocol provably works well when there is one dominant decay parameter that needs to be ascertained. 
When multiple parameters need to be known the difficulty of fitting multiple decay parameters is mitigated by preparing particular states (or applying particular gates) that will maximize one of the pre-factors while minimizing the others (e.g.~\cite{Gambetta2012,Dugas2015,Cross2015,Hashagen2018}) or using projective properties of subgroups of the relevant groups~\cite{Helsen2018}. 
To date, in all cases where multiple parameters need to be found a number of different experiments need to be run to ascertain the fitting parameters. 
Despite the difficulty of this, in practice it has been shown to be successful (e.g.~Ref.~\cite{Harper2018}).

\subsection{Protocol to measure all the single-qubit \texorpdfstring{$f$}{f} parameters simultaneously}\label{sec:singleQubitProtocol}

The idea of the protocol is that rather than using different types of experiments to extract multiple $\lambda$ parameters one at a time, we can use the probability distribution of independent measurements on each qubit to reconstruct the relevant $\{\mathbbm{1},Z\}^{\otimes n}$ $f$ parameters $\lambda_i$, ($i\in\{1..2^n\}$), corresponding to the $2^n$ irreps related to single-qubit Clifford twirls conducted simultaneously on each qubit.
In the next section we will also discuss how to generalize this to the case where we use a multi-qubit Clifford group. 

We can measure all $2^n$ observed error rates by performing an inverse Walsh-Hadamard transform on the $2^n$ different possible $\{\uparrow, \downarrow\}^{\otimes n}$ observation outcomes (see \cref{sec:measurement}) to determine the $\lambda$ parameters that lead to such measurement outcomes. 
For twirled channels, the observed probability distributions will be caused by a combination of both the SPAM and the $\lambda$ parameters associated with the noisy gates. 
We want to be able to isolate the $\lambda$ parameters. 
By estimating the decay parameter (being the combination of both SPAM and gate noise) at varying gate lengths ($m$), we can fit each of these parameters to a formula of the form $Ap(j)^m$, eliminating SPAM noise and (with appropriately chosen $m$), ensuring the error in our estimates are multiplicative in $1-\lambda$ for each of the $\lambda$ being estimated~\cite{Harper2019, Flammia2019}.
Once we have SPAM-free estimates of the $\lambda$ values, we can again use a Walsh-Hadamard transform to reconstruct the SPAM-free estimate of the probability distribution of observed error rates in the machine. 
This has many uses, some of which we detail later.

More precisely, the algorithm for an $n$ qubit system is as follows:

\begin{enumerate}
\item Choose a positive integer $m$.

\item Choose a random sequence of gates $s$ from a set $\mathbb{S}_m$, drawn from the one-qubit Cliffords for each qubit. 
For each qubit sequence, choose the gates randomly to either leave the state of that qubit invariant or to map it on an orthogonal state.

\item Obtain an estimate $\hat{q}(m,s)$ of the probability distribution over the $2^n$ different measurement outcomes for the the $n$ observables (being the measurement of each the $n$ qubits in the device).

\item Repeat steps 2--3 to obtain an estimate $\hat{q}(m)$ of $\bar{q}(m)$, ($\hat{q}(m)$ being a vector with $2^n$ entries one for each of the possible observed outcomes).

\item Transform $\hat{q}(m)$ by applying the Walsh-Hadamard transformation to obtain $\hat{p}(m)$ representing the estimate of the relevant entry in the noise matrix (applied to itself m times).

\item Repeat steps 1--5 and then for each parameter in  $\hat{p}(m)$  fit to the model
\begin{align}
\bar{p}(m) = A p^m 
\end{align}
to obtain SPAM free estimates of the relevant decay parameters. 

\item Use the relevant estimates of the decay parameters (with a forward Walsh Hadamard transform) to reconstruct the entire probability distribution, if necessary projecting onto a simplex to ensure all probabilities are $\geq 0$.
\end{enumerate}

Note that depending on the data required and the size of the system it is perfectly acceptable to marginalize the data for Steps 5 and 6 before fitting, rather than fitting all the data, converting back to probabilities and then marginalizing. 
The Walsh-Hadamard transform and the marginalization of probabilities commute and one can choose whichever route is easier. However, this is not the case with the final projective step onto the probability simplex.

\subsection{Protocol to measure sets of two-qubit  \texorpdfstring{$f$}{f} parameters simultaneously}\label{sec:twoQubitProtocol}

The protocol discussed in \cref{sec:singleQubitProtocol} relied on single-qubit Clifford gates to average the Pauli noise. 
In many architectures the noise on two-qubit gates is quite different from that of single-qubit gate noise. 
Here we present an extension to a protocol that allows the characterization of the noise in the system where such qubit-to-qubit couplings are used.

In order to activate such two-qubit couplings a two-qubit randomized benchmarking protocol can be conducted between qubit pairs. 
The easiest way is just to use two-qubit randomized benchmarking. 
The number of Cliffords to choose from is relatively small (11,520 gates), and can be reduced further using, for example, the group identified in \cite{Cleve2015} (960 gates). 

The protocol is otherwise identical.
However, it should be noted that errors in one qubit will now spread to the other qubit involved in the two-qubit gate, thus their errors will be correlated when the probability distributions are analyzed (see later). 
Finally, when qubits have multiple qubits with which they can interact, the protocol only allows one such link to be tested at a time. 
The two-qubit gates need to be placed into distinct sets. 
If one wishes to analyze every two-qubit link then the number of such sets (and thus the number of times the protocol will need to be repeated) is equal to the number of connections of the most connected qubit, which given current devices will be a manageable number.

It should be noted, however, that the average number of two-qubit gates in a full two-qubit Clifford twirl is not insignificant (using only, say a controlled-$Z$ gate ($\mathcal{C}_Z$), it has an average of approx 1.6, with a maximum of 3) and when all the qubits are active at once this may lead to decay rates that are too large to be measured accurately (i.e.~they decay to the maximally mixed state with just a few gates). 

If one desires to have all the two-qubit gates acting at the same time one could use the protocol given in Ref.~\cite{Helsen2018}, interleaving a two-qubit gate such as the $\mathcal{C}_Z$ gate between two single qubit Clifford twirls. 
However, it should be noted that with low gate numbers, one only gets an approximate exponential decay rate. 
Where we use a Paul twirl instead of a Clifford twirl this non-exponential decay rate does not arise, but then we do not average the Pauli noise, increasing the number of parameters we need to measure.

Alternatively, we can adapt the twirl outlined in~\cite{Harper2017} and perform a $\mathcal{C}\mathcal{C}_Z\mathcal{P}\mathcal{C}_Z$ twirl, where $\mathcal{C}$ is a gate drawn from the group composed of single-Clifford twirls on two qubits, $\mathcal{C}_Z$ is the two-qubit gate and $\mathcal{P}$ is one of the 16 two-qubit Paulis. 
Here we are using the fact that the two-qubit Clifford gate being, by definition, a Clifford gate, conjugates Paulis to Paulis. 
The protocol thus averages all the noise as described before except in this case it is the average of the square of each of the entries of the `noise matrix' rather than just the average of the relative entries.
As discussed in Ref.~\cite{Harper2017} the difference in the value of these figures is related to the anisotropy of the noise. 
Finally we note that the twirls in cycle benchmarking~\cite{Erhard2019} can always be used to ascertain the Pauli noise in the device with interleaved two-qubit gates.

\subsection{Extracting the fidelity}
\label{sec:fidelityExtraction}

It should be noted that the fidelity of the various noise models can be reconstructed trivially from the estimates in step 6 above, recalling that the average gate fidelity is $\mathcal{F} = \frac{(d-1)\bar{\lambda}+1}{d}$, where $\bar{\lambda}$ is the average of the non-identity diagonal elements of the superoperator representing the noise. 
If a set of gates other than the Cliffords is chosen in step 2, then, depending on the group chosen, it will be possible to extract $2^n$ parameters of interest per run. 
For instance, the real Cliffords \cite{Hashagen2018} (of interest, for example, for codes that might not have a fault-tolerant phase gate \cite{Harper2018}) when used as the twirling group gives three decay parameters. 
With the correct preliminary state (or initial gates) all decay parameters reside on one of the $\lambda_{\{I,Z\}}$ parameters extracted, allowing the fidelity to be calculated without the need for multiple different types of experiments.

Because of the ability to transform the decay parameters (the $\lambda$ values) to a probability distribution (and vice-versa), when only subsets of the qubits are of interest the relevant decay parameters for such subsets are easily obtainable by converting to a probability distribution, marginalizing, and then converting back.

\begin{figure}[t!]
    \centering
    \includegraphics[width=0.48\textwidth]{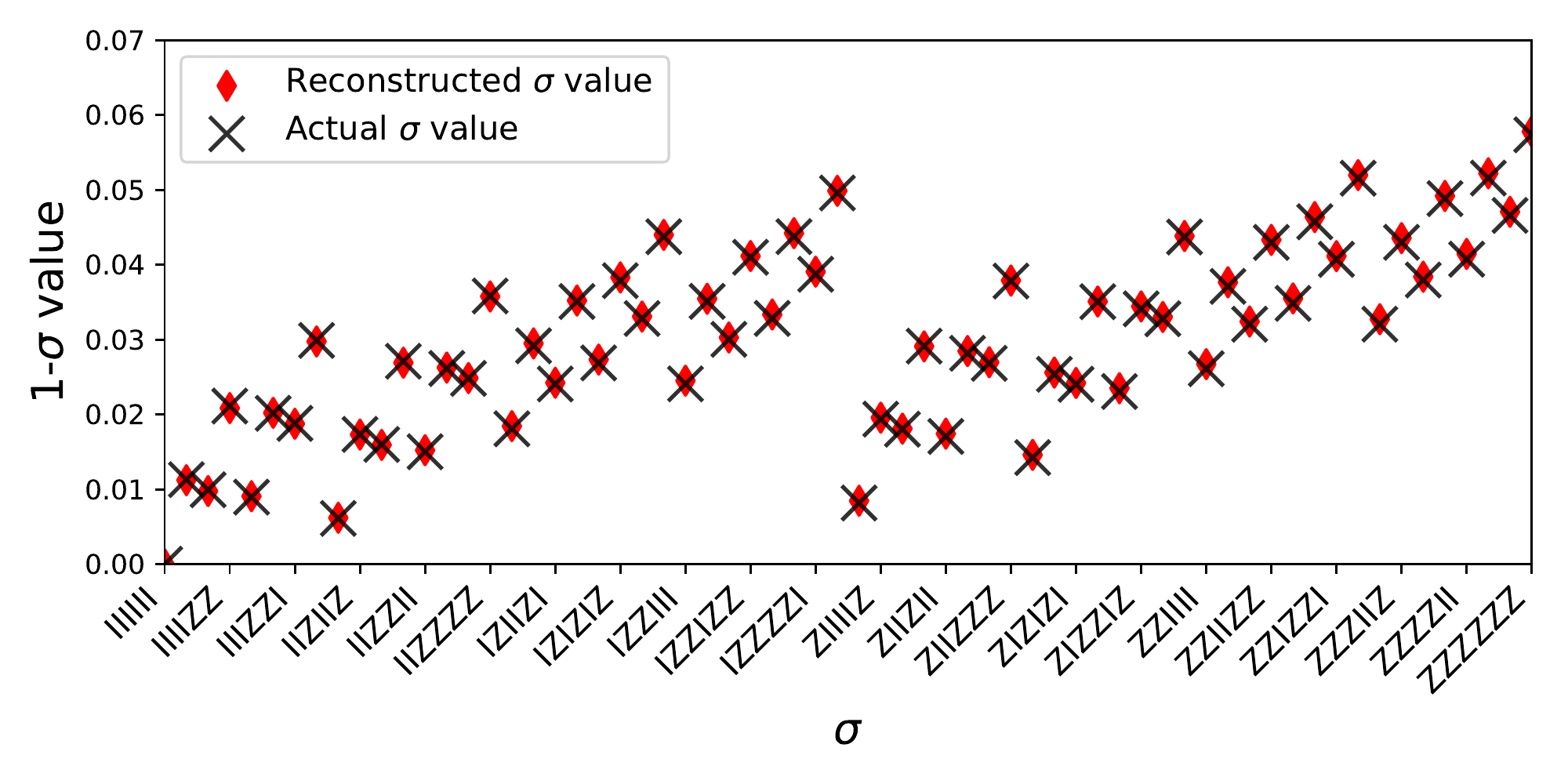}
    \caption{Numerical simulation with a reconstruction of the 64 averaged elements of the 6 qubit noise matrix on the simulated system. 
    Even with relatively few sequences (50) over just 11 different gate lengths, the relative error in any of the reconstructed 64 entries is less than 2\%.}
    \label{fig:reconstructed}
\end{figure}

\begin{figure*}[t!]
    \includegraphics[width=1\textwidth]{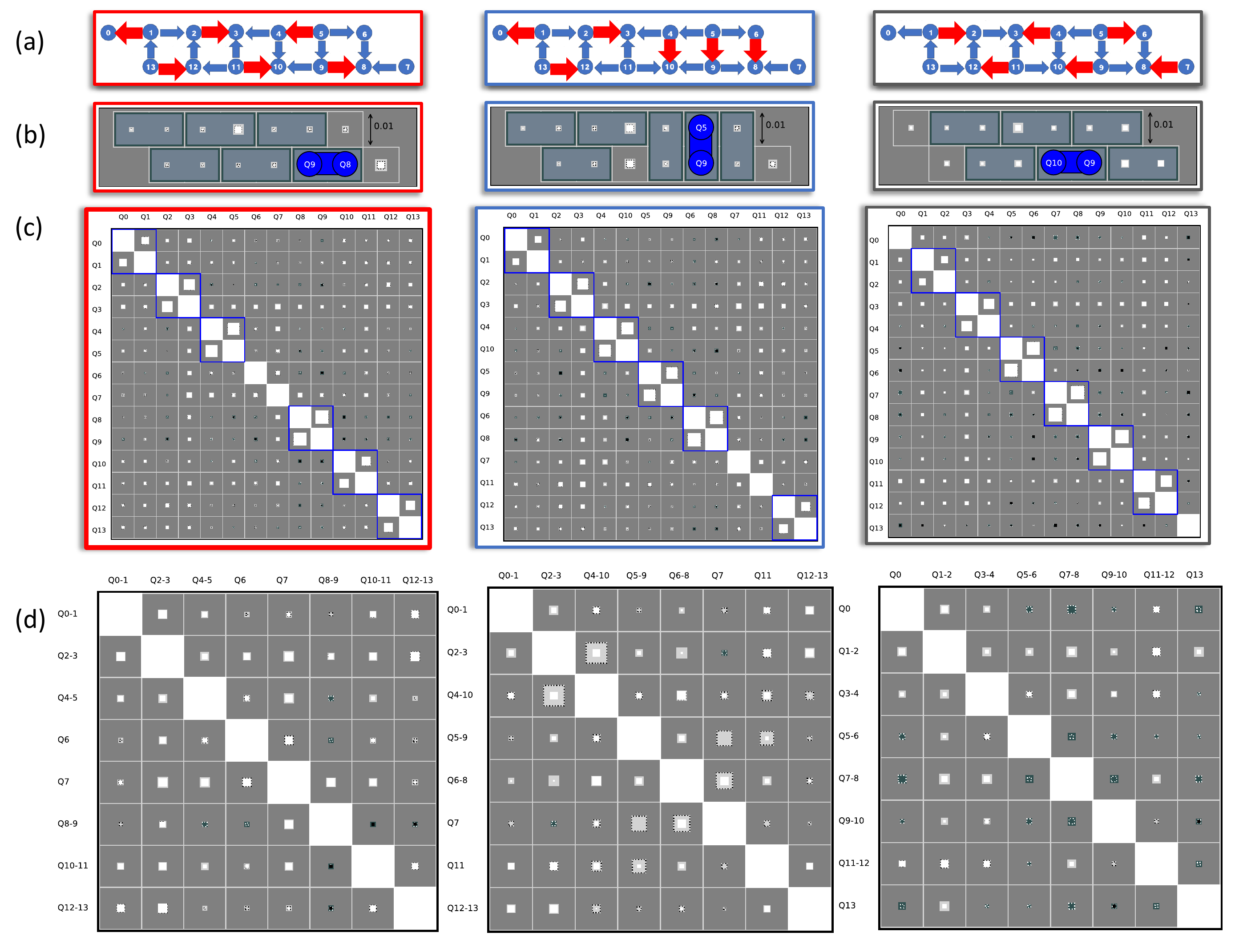}
    \caption{(a) Schematic showing the activated two-qubit gates (in red) for each of the three protocols. 
    (b) Hinton diagrams showing the mutual information between the highlighted qubit pairs and each of the other qubits. 
    The darker shading in the diagrams indicates which qubits were operated in two qubit mode. 
    The area of the white square indicates the strength of the mutual information. 
    (c) The covariance matrices for each of the three protocols. 
    The qubits that formed part of the two-qubit Clifford twirl are outlined in blue. 
    Note in the second column the order of the qubits has been changed to ensure that qubits that are part of a two-qubit twirl are next to each other. 
    (d) The covariance matrix where each of the qubits forming part of a two-qubit twirl are treated as a single random variable. 
    This allows the correlations between these qubit pairs and the other qubit pairs to be seen more clearly.}
    \label{fig:twoQreconstructed}
\end{figure*}

\section{Meaning of ``Observed Error Rate''}

In the main paper we say:

\vspace{12pt}
\hrule
\begin{quote}
This procedure provably converges to an estimate of the probability
distribution of the average noise in the
system \cite{Flammia2019}, though the variant
implemented here uses random gates chosen from the single-qubit Clifford
group instead of the Pauli group. This leads to a simpler algorithm, but
one that additionally averages over the local basis information.
Averaging the noise in this way reduces the number of parameters that
need to be reconstructed from \(4^n\) to \(2^n\). We call this reduced
distribution the observed error rates to contrast with the larger
distribution of Pauli error rates, and we similarly define the set of
\(2^n\) averaged eigenvalues in analogy with the \(4^n\) eigenvalues.
Importantly, the observed error rates are still capable of describing all
many-body correlations in the noise.
\end{quote}
\hrule
\vspace{12pt}

Here we wish to expand on this claim to make the terms used explicit.

There are three relevant parts to the paragraph quoted above:

\begin{enumerate}

\item
  What are the observed error rates?
\item
  Are the observed error rates a meaningful quantity?
\item
  Are the observed error rates still sufficient to describe all many-body
  correlations in the noise?
\end{enumerate}

\subsection{What is the definition of the observed error rates?}

The idea  is that we we have the $2^n$ distribution delivered as per the algorithm in Figure~1 of the main paper. We call this reduced probability distribution the ``observed error" rates. In this case, analogous to the Pauli error rates, it is a joint distribution showing the chance of seeing a particular error pattern of errors on the qubits. Not to be confused with the chance of an error on a particular qubit, although of course that can be determined by marginalising the joint probability distribution appropriately.

It is instructive to compare these observed error rates, with their analogous Pauli error rates. Here are some arbitrary numbers for a Pauli twirled single qubit channel, with \(4^n\) numbers e.g.

 \begin{equation} \left(\begin{array}{*{11}c} 1.0 & 0.0 & 0.0 & 0.0 & \\ 0.0 & 0.92 & 0.0 & 0.0 & \\ 0.0 & 0.0 & 0.94 & 0.0 & \\ 0.0 & 0.0 & 0.0 & 0.96 &  \end{array}\right)\\\end{equation}

(Where we are using the following basis I, X, Y, Z). Because we are twirling the channel (either Pauli or Clifford) we are now only interested in the diagonal elements.
 
We can use the Walsh-Hadamard transform to  transform this into a probability distribution

\begin{equation} \left(\begin{array}{*{1}c} 0.955 \\ 0.015 \\ 0.005  \\ 0.025  \end{array}\right)\\\end{equation}

And this tells us the chance of no error on the qubit is 0.955, the chance of an X error on the qubit is 0.015, Y error .005 and Z error 0.025. This is the full Pauli error rates distribution we are contrasting with.

Now what about our ``averaged" protocol, well we do a Clifford twirl:

   \begin{equation} \left(\begin{array}{*{11}c} 1.0 & 0.0 & 0.0 & 0.0 & \\ 0.0 & 0.94 & 0.0 & 0.0 & \\ 0.0 & 0.0 & 0.94 & 0.0 & \\ 0.0 & 0.0 & 0.0 & 0.94 &  \label{eq:rawPauli}\end{array}\right)\\\end{equation}

But then, rather than doing a transform on this to get:

\begin{equation} \left(\begin{array}{*{1}c} 0.955 \\ 0.015 \\ 0.015  \\ 0.015  \end{array}\right)\label{eq:averagePauli}\end{equation}

(which would tell us that the chance of no error on the qubit is 0.955, the chance of an X error on the qubit is 0.015, Y error .015 and Z error 0.015) what we do is we only look at the reduced 'qubit eigenvalues'

\begin{equation} \left(\begin{array}{*{1}c} 0.955 \\ 0.015  \end{array}\right)\\\label{eq:singleCompressed}\end{equation}

Do the $2^n\times2^n$ transform, to get the \textbf{observed error rates}:

\begin{equation} \left(\begin{array}{*{1}c} 0.97 \\ 0.03  \end{array}\right)\\\end{equation}

And it is this distribution $\{0.97, 0.03\}$ that constitutes the ``observed error rates''. Obviously with one qubit it is not much of a distribution, but we make this clearer when we consider two qubits. It is worth noting that this is not the same as  the sum of the ``Pauli-error rates''. 

If one wanted to get the entire Pauli channel without the averaging, then ref~\cite{Flammia2019} contains the requisite algorithms.

\subsection{Are the observed error rates a meaningful quantity?}

The observed error rates as defined only make intuitive sense when we have Clifford averaged the noise (this might be clearer when we look at two qubits). The point is that we have ``locally averaged the noise"

Given this the first thing to note is that we can take that distribution 

\begin{align}
    \text{p(success) }&= 0.97\\
    \text{p(fail)} &= 0.03 
\end{align}

and recreate the matrix in \cref{eq:averagePauli}. 
A Walsh-Hadamard transform recreates \cref{eq:singleCompressed} and then it is trivial to reconstruct \cref{eq:averagePauli}. For the sake of completeness we discuss how to do this for a multi qubit system at the end of this section. Obviously we cannot recreate the matrix in \cref{eq:rawPauli}, we have lost the information required to do that. 
However we \emph{can}  recreate \cref{eq:averagePauli}. 

So how these different probabilities related, well quite simply:

\begin{equation}
    0.97 = 0.955 + 1*(0.015+0.015+0.015)/3
\end{equation}
and 
\begin{equation}
0.03 = 2*(0.015+0.015+0.015)/3\label{eq:singleFormula}
\end{equation}

And this gives us our operational insight. We are observing only in the computational basis state and the $2^n$ transform gives us the chance of ``observing" an error on the qubit.

If a Pauli error occurs, then $\frac{2}{3}$'s of the time, it is Clifford twirled to a basis where the error can be observed, and $\frac{1}{3}$ of the time it is not observed. Of course averaged over the circuits, it means we will detect the combined Pauli errors $\frac{2}{3}$'s of the time, giving rise to \cref{eq:singleFormula}.

However, if we ever needed to recreate the averaged eigenvalues or indeed the Pauli error distribution, we have all of the information we need --- if we allow the assumption that we have correctly Clifford twirled the noise then this observation and observed error distribution allows us to recreate \cref{eq:averagePauli}.

So the question: ``Are the observed error rates a meaningful quantity" - is answered in the affirmative. They correspond directly the chance of ``seeing" an error pattern on the qubits and in any case (under the assumption of Clifford twirl) they allow us to reconstruct the full (averaged but not reduced) Pauli noise matrix.

In this very small probability distribution, one could also think of it as

\begin{align}
p(\text{no error observed}) = 0.97 \nonumber\\
p(\text{error observed}) = 0.03
\end{align}

Which we can compare with the ``averaged Pauli global distribution"
\begin{align}
p(\text{no error}) = 0.955\nonumber\\
p(\text{x error}) = 0.015\nonumber\\
p(\text{y error}) = 0.015\nonumber\\
p(\text{z error}) = 0.015
\end{align}

One could say observed error rates are ``informationally interchangeable" with the ``averaged global Pauli distribution'', for the reasons described above.

\subsection{Are the observed error rates are still capable of describing all many-body correlations in the noise?}

To explore this we are going to have to go to two qubits. What will quickly become  clear is just how much more compact the reduced probability distribution is.

The outline is:

\begin{itemize}

\item
  to create a two qubit depolarising channel (i.e.~Pauli twirled) and look at how the full transform works,
\item
  single Clifford twirl it, and look at the how the two different algorithms (\cite{Flammia2019} and this paper) work
\item
  check the qubit error rates are not correlated
\item
  finally introduce an arbitrary correlation on two Paulis (we choose
  XY but it works for any Paulis including, say, ZZ) and
  show how the averaging algorithm still picks that out.
\end{itemize}

    So we choose some arbitrary numbers to give us slightly different errors on the second qubit. These errors are still independent at this stage. 

On qubit 0

    \begin{equation} \left(\begin{array}{*{11}c} 1.0 & 0.0 & 0.0 & 0.0 & \\ 0.0 & 0.92 & 0.0 & 0.0 & \\ 0.0 & 0.0 & 0.94 & 0.0 & \\ 0.0 & 0.0 & 0.0 & 0.96 & \\ \end{array}\right)\\\end{equation}

And on qubit 1

  \begin{equation} \left(\begin{array}{*{11}c} 1.0 & 0.0 & 0.0 & 0.0 & \\ 0.0 & 0.95 & 0.0 & 0.0 & \\ 0.0 & 0.0 & 0.93 & 0.0 & \\ 0.0 & 0.0 & 0.0 & 0.91 & \\ \end{array}\right)\\\end{equation}

If we tensor the qubits together we note the following eigenvalues, and the Walsh-Hadamard transform gives us the probability distribution shown in \cref{tab:i2qe}.
\begin{center}
\begin{longtable}{|l|l|l|}
\caption{Independent 2 qubit pauli errors.} \label{tab:i2qe} \\

\hline \multicolumn{1}{|c|}{\textbf{Pauli}} & \multicolumn{1}{c|}{\textbf{Eigenvalue}} & \multicolumn{1}{c|}{\textbf{Probability}} \\ \hline 
\endfirsthead

\multicolumn{3}{c}%
{{\bfseries \tablename\ \thetable{} -- continued from previous page}} \\
\hline \multicolumn{1}{|c|}{\textbf{Pauli}} & \multicolumn{1}{c|}{\textbf{Eigenvalue}} & \multicolumn{1}{c|}{\textbf{Probability}} \\ \hline 
\endhead

\hline \multicolumn{3}{|r|}{{Continued on next page}} \\ \hline
\endfoot

\hline \hline
\endlastfoot

$II$& 1.0& 0.9048625\\
$IX$& 0.95& 0.0167125\\
$IY$& 0.93& 0.0262625\\
$IZ$& 0.91& 0.0071625\\
$XI$& 0.92& 0.0142125\\
$XX$& 0.874& 0.0002625\\
$XY$& 0.8556& 0.0004125\\
$XZ$& 0.8372& 0.0001125\\
$YI$& 0.94& 0.0047375\\
$YX$& 0.893& 8.75e-5\\
$YY$& 0.8742& 0.0001375\\
$YZ$& 0.8554& 3.75e-5\\
$ZI$& 0.96& 0.0236875\\
$ZX$& 0.912& 0.0004375\\
$ZY$& 0.8928& 0.0006875\\
$ZZ$& 0.8736& 0.0001875\\

\end{longtable}
\end{center}

So we can twirl this (single Clifford twirls) with the results shown in \cref{tab:t2qe}.

\begin{center}
\begin{longtable}{|l|l|l|}
\caption{Independent 2 qubit pauli errors- clifford twirled.} \label{tab:t2qe} \\

\hline \multicolumn{1}{|c|}{\textbf{Pauli}} & \multicolumn{1}{c|}{\textbf{Eigenvalue}} & \multicolumn{1}{c|}{\textbf{Probability}} \\ \hline 
\endfirsthead

\multicolumn{3}{c}%
{{\bfseries \tablename\ \thetable{} -- continued from previous page}} \\
\hline \multicolumn{1}{|c|}{\textbf{Pauli}} & \multicolumn{1}{c|}{\textbf{Eigenvalue}} & \multicolumn{1}{c|}{\textbf{Probability}} \\ \hline 
\endhead

\hline \multicolumn{3}{|r|}{{Continued on next page}} \\ \hline
\endfoot

\hline \hline
\endlastfoot

$II$&1.0& 0.9048625\\
$IX$&0.93& 0.0167125\\
$IY$&0.93& 0.0167125\\
$IZ$&0.93& 0.0167125\\
$XI$&0.94& 0.0142125\\
$XX$&0.8742& 0.0002625\\
$XY$&0.8742& 0.0002625\\
$XZ$&0.8742& 0.0002625\\
$YI$&0.94& 0.0142125\\
$YX$&0.8742& 0.0002625\\
$YY$&0.8742& 0.0002625\\
$YZ$&0.8742& 0.0002625\\
$ZI$&0.94& 0.0142125\\
$ZX$&0.8742& 0.0002625\\
$ZY$&0.8742& 0.0002625\\
$ZZ$&0.8742& 0.0002625\\

\end{longtable}
\end{center}

This is exactly as expected - and the reduced algorithm (i.e. the one if figure 1) gives us the data shown in \cref{tab:r2qe}

\begin{center}
\begin{longtable}{|l|l|l|l|}
\caption{Indep. 2 qubit Pauli errors- Clifford twirled and reduced.} \label{tab:r2qe} \\

\hline \multicolumn{1}{|c|}{\textbf{Label}}&\textbf{Obs.} & \multicolumn{1}{c|}{\textbf{Eigenvalue}} & \multicolumn{1}{c|}{\textbf{Probability}} \\ \hline 
\endfirsthead

\multicolumn{3}{c}%
{{\bfseries \tablename\ \thetable{} -- continued from previous page}} \\
\hline \multicolumn{1}{|c|}{\textbf{Label}}&\textbf{Obs.} & 
\multicolumn{1}{c|}{\textbf{Probability}} \\ \hline 
\endhead

\hline \multicolumn{3}{|r|}{{Continued on next page}} \\ \hline
\endfoot

\hline \hline
\endlastfoot

${II}$&$\uparrow\uparrow$&1.0& 0.93605\\
$I\Sigma$&$\uparrow\downarrow$&0.93& 0.03395\\
$\Sigma I$ &$\downarrow\uparrow$&0.94& 0.02895\\
$\Sigma\Sigma$ &$\downarrow\downarrow$&0.8742& 0.00105\\

\end{longtable}
\end{center}

and if we generate the `mini-correlation matrix' then there are no correlations as we expected

    \begin{equation} \left(\begin{array}{*{11}c} 1.0 & 0.0 & \\ -0.0 & 1.0 & \\ \end{array}\right)\\\end{equation}

Just for completeness we can tell from the original probability distribution that quite clearly everything is independent:

\begin{center}
\begin{longtable}{|l|l|l|l|}
\caption{Demonstration of Indep.} \label{tab:t2qe2} \\

\hline Q0 &Q1&Q0*Q1&From \cref{tab:i2qe} \\ \hline 
\endfirsthead

\multicolumn{3}{c}%
{{\bfseries \tablename\ \thetable{} -- continued from previous page}} \\
\hline Q0 &Q1&Q0*Q1&From \cref{tab:i2qe} \\ \hline 
\endhead

\hline \multicolumn{3}{|r|}{{Continued on next page}} \\ \hline
\endfoot

\hline \hline
\endlastfoot

p(I)=0.955 &  p(I)=0.9475&p(II)=0.9048625& 0.9048625\\
p(I)=0.955 &  p(X)=0.0175&p(IX)=0.0167125& 0.0167125\\
p(I)=0.955 &  p(Y)=0.0275&p(IY)=0.0262625& 0.0262625\\
p(I)=0.955 &  p(Z)=0.0075&p(IZ)=0.0071625& 0.0071625\\
p(X)=0.015 &  p(I)=0.9475&p(XI)=0.0142125& 0.0142125\\
p(X)=0.015 &  p(X)=0.0175&p(XX)=0.0002625& 0.0002625\\
p(X)=0.015 &  p(Y)=0.0275&p(XY)=0.0004125& 0.0004125\\
p(X)=0.015 &  p(Z)=0.0075&p(XZ)=0.0001125& 0.0001125\\
p(Y)=0.005 &  p(I)=0.9475&p(YI)=0.0047375& 0.0047375\\
p(Y)=0.005 &  p(X)=0.0175&p(YX)=8.75e-5  & 8.75e-5\\
p(Y)=0.005 &  p(Y)=0.0275&p(YY)=0.0001375& 0.0001375\\
p(Y)=0.005 &  p(Z)=0.0075&p(YZ)=3.75e-5  & 3.75e-5\\
p(Z)=0.025 &  p(I)=0.9475&p(ZI)=0.0236875& 0.0236875\\
p(Z)=0.025 &  p(X)=0.0175&p(ZX)=0.0004375& 0.0004375\\
p(Z)=0.025 &  p(Y)=0.0275&p(ZY)=0.0006875& 0.0006875\\
p(Z)=0.025 &  p(Z)=0.0075&p(ZZ)=0.0001875& 0.0001875\\
\end{longtable}
\end{center}

Let's see what happens if we add an XY correlation, i.e more likely to see X error on Q0 if we have a Y error on Q1 (or vice-versa). (The Pauli we choose here is irrelevant).

So we change XY from 0.0004125, so say 0.0006. In this case we renormalise by `stealing' probability from II, so we have a new probability distribution. (This is just for demonstration, the numerics in \cref{sec:numerics} use simulated correlated noise.) \Cref{tab:2qcorr} shows the resulting Pauli eigenvalues.

\begin{center}
\begin{longtable}{|l|l|l|l|}
\caption{Pauli algorithm with XY correlation added.} \label{tab:2qcorr} \\

\hline \multicolumn{1}{|c|}{\textbf{Pauli}} & \textbf{Altered dist.} & \multicolumn{1}{c|}{\textbf{Transform to  $\lambda$}} & \multicolumn{1}{c|}{\textbf{twirled $\lambda$}} \\ \hline 
\endfirsthead

\multicolumn{3}{c}%
{{\bfseries \tablename\ \thetable{} -- continued from previous page}} \\
\hline \multicolumn{1}{|c|}{\textbf{Pauli}}  & \textbf{altered distribution}& \multicolumn{1}{c|}{\textbf{Eigenvalue}} & \multicolumn{1}{c|}{\textbf{twirled $\lambda$}} \\ \hline 
\endhead

\hline \multicolumn{3}{|r|}{{Continued on next page}} \\ \hline
\endfoot

\hline \hline
\endlastfoot

$II$&\textbf{0.899275}& 1.0& 1.0\\
$IX$&0.0167125& 0.95& 0.92255\\
$IY$&0.0262625& 0.918825& 0.92255\\
$IZ$&0.0071625& 0.898825& 0.92255\\
$XI$&0.0142125& 0.908825& 0.93255\\
$XX$&0.0002625& 0.862825& 0.86923333\\
$XY$&\textbf{0.006}& 0.8556& 0.86923333\\
$XZ$&0.0001125& 0.8372& 0.86923333\\
$YI$&0.0047375& 0.94& 0.93255\\
$YX$&8.75e-5& 0.893& 0.86923333\\
$YY$&0.0001375& 0.863025& 0.86923333\\
$YZ$&3.75e-5& 0.844225& 0.86923333\\
$ZI$&0.0236875& 0.948825& 0.93255\\
$ZX$&0.0004375& 0.900825& 0.86923333\\
$ZY$&0.0006875& 0.8928& 0.86923333\\
$ZZ$&0.0001875& 0.8736& 0.86923333\\

\end{longtable}
\end{center}

We can now use the figures in \cref{tab:2qcorr} to recreate the results we would get if such noise existed in the device and we had followed the steps shown in Figure 1 of the main paper. \Cref{tab:cmini} sets out the relevant qubit eigenvalues, the observations they relate to, the eigenvalues that would come from the fitting procedure and the observed error distribution that results from the final Walsh-Hadamard transform. (For the purposes of this illustration, we have just copied over the eigenvalues from \cref{tab:2qcorr}. In \cref{sec:numerics} we simulate the entire process.)

\begin{center}
\begin{longtable}{|l|l|l|l|}
\caption{Altered 2 qubit pauli errors- clifford twirled and reduced.} \label{tab:cmini} \\

\hline \multicolumn{1}{|c|}{\textbf{Label}}&\textbf{Obs.} & \multicolumn{1}{c|}{\textbf{Eigenvalue}} & \multicolumn{1}{c|}{\textbf{Probability}} \\ \hline 
\endfirsthead

\multicolumn{3}{c}%
{{\bfseries \tablename\ \thetable{} -- continued from previous page}} \\
\hline \multicolumn{1}{|c|}{\textbf{Label}}&\textbf{Obs.} & 
\multicolumn{1}{c|}{\textbf{Probability}} \\ \hline 
\endhead

\hline \multicolumn{3}{|r|}{{Continued on next page}} \\ \hline
\endfoot

\hline \hline
\endlastfoot

${II}$&$\uparrow\uparrow$&1.0& 0.93108333\\
${I\Sigma}$&$\uparrow\downarrow$&0.92255& 0.03519167\\
${\Sigma I}$&$\downarrow\uparrow$&0.93255& 0.03019167\\
${\Sigma\Sigma}$&$\downarrow\downarrow$&0.86923333&  0.00353333\\

\end{longtable}
\end{center}

When we use the probability distribution of \cref{tab:cmini} to create a correlation matrix we get the following:

 \begin{equation*} \left(\begin{array}{*{11}c} 1.0 & 0.0639497067 & \\ 0.0639497067 & 1.0 & \\ \end{array}\right)\\\end{equation*}

The non-zero off diagonal elements show us that the correlation has been picked up.

\subsection{Converting between twirled Pauli eigenvalues/Pauli error rates and qubit eigenvalues}\label{sec:converting}

The final part of this section is to describe how to move from the equivalent of the qubit error rates shown in  \cref{tab:cmini} to the column headed \textbf{twirled $\lambda$} in \cref{tab:2qcorr}. 
At that point all the procedures in Ref.~\cite{Flammia2019} are immediately applicable.

One can see from the way we have labelled the qubit eigenvalues in \cref{tab:cmini} that the patterns of $I$ and $\Sigma$ are just appropriately labelled bit patterns. 
For instance if we had a distribution of $2^8$ values, then there would be 256 probabilities, and then, say, the 123rd one would have a bit pattern of $01111011$. 
This corresponds to a qubit eigenvalue of $\lambda_{I\Sigma\Sigma\Sigma\Sigma I\Sigma\Sigma}$. 
Every Pauli that follows this pattern (where the $I$ requires the trivial Pauli on that qubit and the $\Sigma$ admits any of the $X$, $Y$ or $Z$ Paulis) shares the same eigenvalue as this qubit eigenvalue. 
Given this we can immediately see, for instance, that the Pauli eigenvalue for $IXYZYIZZ$ would have the same value as the 123rd entry of the qubit eigenvalue matrix (which is the Walsh-Hadamard transform of the observed error probabilities). 
Consequently to fill in a table for Pauli eigenvalues, replace each $I$ with 0, and every $X$,$Y$ and $Z$ with 1, interpret this as a binary value and use it to index the qubit eigenvalue vector.

More formally,
let $\bs{\lambda}$ and $\bs{\mu}$ be the vectors of Pauli channel eigenvalues and Pauli error rates ordered lexographically to be consistent with the tensor product.
From \cite{Flammia2019}, we have
\begin{align}
    \bs{\lambda} = W^{\otimes n} \bs{\mu}
\end{align}
where
\begin{align}
    W = \begin{pmatrix}
    1 & 1 & 1 & 1 \\
    1 & 1 & -1 & -1 \\
    1 & -1 & 1 & -1 \\
    1 & -1 & -1 & 1
    \end{pmatrix}.
\end{align}
Define the reduced vector of eigenvalues $\bs{\lambda}_\Sigma$ by removing all eigenvalues with $X, Y$, so that under perfect Clifford averaging,
\begin{align}
    \bs{\lambda} = (M^T)^{\otimes n} \bs{\lambda}_\Sigma
\end{align}
where
\begin{align}
    M = \begin{pmatrix}
    1 & 0 & 0 & 0\\
    0 & 1 & 1 & 1
    \end{pmatrix}.
\end{align}
Then the full vector of Pauli error rates can be written in terms of the reduced vector of eigenalues as
\begin{align}
    \bs{\mu} = (W^{-1} M^T)^{\otimes n} \bs{\lambda}_\Sigma.
\end{align}
To convert back to the averaged error rates with $\bar{p}$, we have
\begin{align}
    \bs{\mu}_\Sigma = (M W^{-1} M^T)^{\otimes n} \bs{\lambda}_\Sigma = N^{\otimes n} \bs{\lambda}_\Sigma
\end{align}
where
\begin{align}
    N = \frac{1}{4}\begin{pmatrix}
    1 & 3 \\
    3 &-3 
    \end{pmatrix}.
\end{align}

Finally we can convert between the vector of averaged error rates ($\bs{\mu}_\Sigma$) and the observed error rates ($\mu_o$) by noting that $H^{\otimes n}\mu_o=\bs{\lambda}_\Sigma$ where
\begin{align}
    H = \begin{pmatrix}
    1 & 1 \\ 1 & -1
    \end{pmatrix}.
\end{align}
Therefore
\begin{align}
    \bs{\mu}_\Sigma = (NH)^{\otimes n} \mu_o
\end{align}

\begin{figure*}
	\centering
	\begin{tikzpicture}
	\node at (0,0) {\includegraphics[width=1\textwidth]{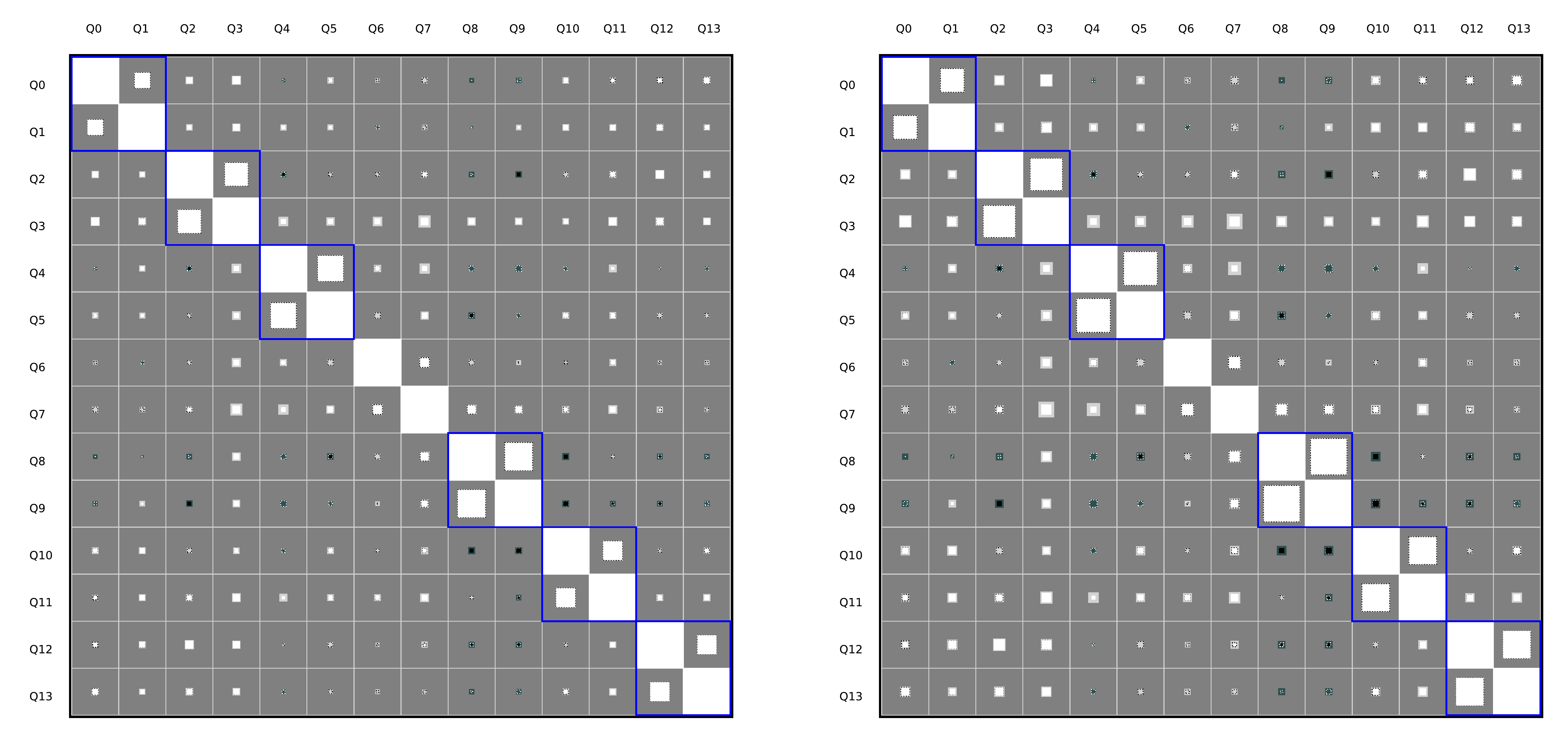}};
	\node at (-8.5,4) {(a)};
	\node at (0.5,4) {(b)};
	
	\end{tikzpicture}
	\caption{Comparison between correlation matrices produced using the observed error rates and the averaged Pauli error rates. Sub-figure (a) shows the correlation matrix constructed from observed error rates, sub-figure (b) the one from averaged Pauli error rates (see \cref{sec:converting}).  The pattern is identical confirming that the observed error rate distribution preserves the correlations between errors on the qubits, although the magnitude of the correlations are decreased with observed error rates, as compared to the Pauli error rates. Since any particular correlation on the figure is as between two particular qubits, the data can be marginalised to determine, with high accuracy, the magnitude difference between any particular values in the charts. }\label{fig:comparison}
\end{figure*}

Using this relationship, we can examine the difference in the correlation matrices obtained from the two distributions (\cref{fig:comparison}). 
As might be anticipated, the absolute value of the Pauli error rate correlations are larger than the observed error rates, but the correlation patterns are the same.

\section{Scalable Estimations}

As discussed in the main text, as the number of qubits being considered increases it will no longer be feasible to reconstruct the entire probability distribution of observed error rates or Pauli error rates (the probability distribution itself growing exponentially with the number of qubits). 
However, reconstruction of covariance matrices and, more importantly, a Gibbs random field (GRF) decomposition, will remain scalable and can be conducted in polynomial time. 
The Walsh-Hadamard transform commutes with marginalizing the data, thus we can marginalize at the experiment level, transform, fit the data, and transform back to compute all relevant marginal joint probability distributions.

Using this technique will allow all covariances or correlations between the qubits to be determined in scalable way. 
This holds similarly for the case of Gibbs random field decompositions, where such decompositions will be motivated by the physical layout of the device.

In the presence of correlations, the probability of, e.g., an $X$ error on qubit 1 and no error on qubit $j$ will depend on which other qubit $j$ is being considered.
The estimation of a globally consistent GRF and a bound on the error incurred at this step have been bounded in~\cite{Flammia2019}. 
A similar technique can be used with the marginalised data to determine the parameters of the GRF decomposition being used to model the underlying probability distribution. 
Thus within the ansatz of the GRF decomposition the probability distribution can be learned within polynomial time up to an overall normalizing factor, assuming that the GRF has bounded degree.

In the present manuscript, the $2^{14}$ values required for estimation of the observed error rates are easily manipulated by current computers and the reconstructed probability distribution (with possible negative values) was projected onto the nearest (in Euclidean distance) point in the probability simplex. 
When such a brute-force manipulation is not possible, we need to determine the best consistent estimator for the reconstruction of the global covariance matrix from the (possibly non-consistent) marginalized covariance matrices. 

It is an open question what is the best method of covariance matrix estimation in the context of learning SPAM-free observed error rates or Pauli error rates. 
It is also open if methods analogous to the case of matrix product state learning~\cite{Cramer2010a, Lanyon2017} can be generalized to quantum channels in a provable way. 
Such a result would be quite analogous to the GRF decomposition, but would allow for the modeling of more general quantum noise sources.

\section{Running the protocol in simulation and on a device}\label{sec:numerics}

For the purposes of demonstrating the effectiveness of the protocol we simulated a six-qubit system with two different styles of noise and both the single-qubit protocol (\autoref{sec:singleQubitProtocol}) and the two-qubit protocol (\autoref{sec:twoQubitProtocol}).
The noise models are described in some detail in the caption of \autoref{fig:simulation}: we generate arbitrary high-fidelity single qubit noise using the techniques described in \cite{Rudnicki2018}, and two-qubit noise is modeled as additional small controlled-qubit rotations between relevant qubits.

In the single-qubit protocol the noise is simply modeled as a superoperator noise matrix applied after each application of the six single-qubit gates. 
The noise matrix in the superoperator basis is, in this case,  a $4^{6}\times4^{6}$ matrix. 
The protocol allows the extraction of the $2^6$ averaged $\lambda$ values from the total $2^{12}$ diagonal values comprised in the noise matrix and the reconstruction of the actual probability distributions over the $2^6$ possible measurement outcomes. 
For this the simulation sequence lengths of 1, 3 , 5 up to 22 gates were measured, with each length having 50 different sequences, 8096 shots being taken for each sequence. 
As can be seen in \autoref{fig:reconstructed} even with this limited number of sequences this allowed an accurate reconstruction of the diagonal elements of the noise matrix (the relative error in each element was no more than, and often much less than, 2\%). 
This would allow the fidelities of any of the qubits or collection of qubits to be read off with high accuracy. 
The pattern of infidelity values in the chart reflects the higher noise on qubits 2 and 5.

Similarly with the two-qubit protocol, save that there will only be about $2^{\frac{n}{2}}$ values to extract since the two-qubit twirl only has two irreps over a pair of qubits, the respective entries in the reconstructed noise map could be averaged for increased accuracy.

\section{Two-qubit protocol on the 14-qubit architecture}

As discussed in the main paper, the two-qubit protocol was executed for three different patterns of activation of two qubit gates.
\autoref{fig:twoQreconstructed} shows the results of all three experiments and the correlation matrices extracted from the obtained data. 
Strikingly, it can be seen that in all cases qubit 3 became strongly correlated with all the other qubits in the machine once the two-qubit gates (and the time delays related to the use of two-qubit gates) were implemented. 
As discussed in the main paper, the JSD between the probability distribution measured using the the two-qubit protocol and the local GRF model has increased by almost an order of magnitude. 
For each of the layouts shown they are: $0.216(1)$, $0.218(3)$ and $0.212(2)$, respectively. Finally we note that in the correlation matrices there are some differences in the correlations between qubits (j,l) and (k,l) where (j,k) are operated as a two-qubit twirl. 
This implies that in such cases decay rates that should be identical because of the two-qubit twirl are empirically distinct. 
While the cause for this is left for future work, it is interesting that the protocol is able to highlight such matters.

\subsection{Further experiments relating to the two-qubit protocol}

\begin{figure*}
	\centering
	\begin{tikzpicture}
	\node at (-13,4) {\includegraphics[width=0.9\textwidth]{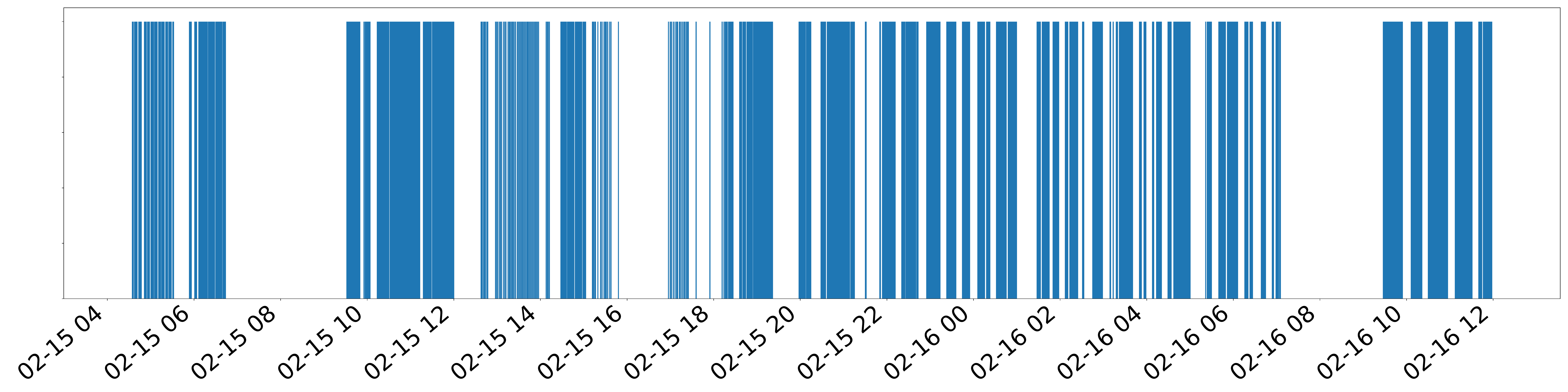}};
	\node at (-13,1) {(a)};
	\node at (-13,-2) {\includegraphics[width=0.9\textwidth]{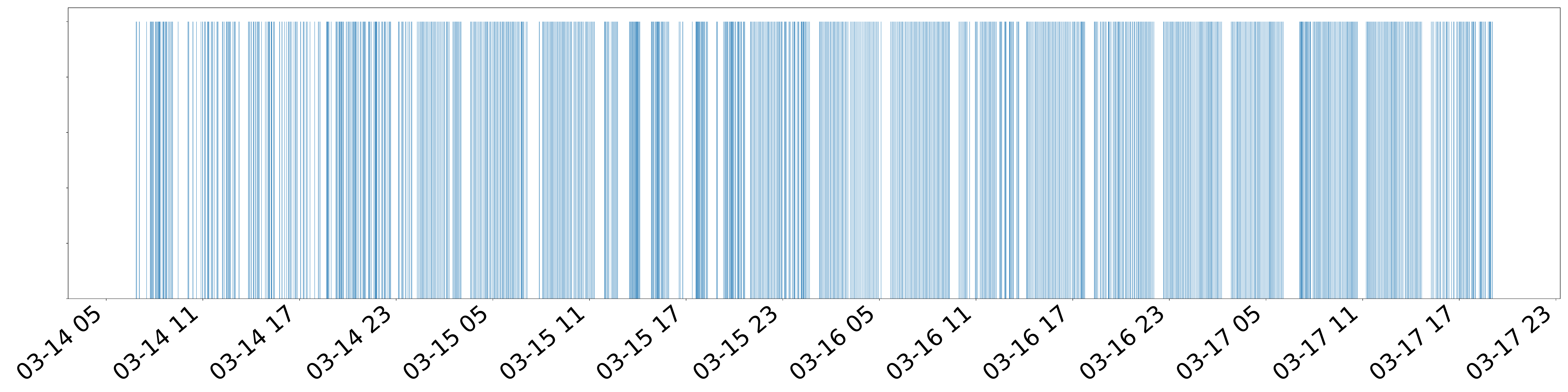}};
	\node at (-13,-5) {(b)};
	\end{tikzpicture}
\caption{Time stamps of pickles with the results of the runs for the two different experiments. (a) shows the single qubit protocol experiment.(1000 pickles) (b) shows the two-qubit experiment. (973 pickles)}\label{fig:timing}
\end{figure*}

Having analysed the data presented in this paper it was realised that on the device in question, two qubit gates took substantially more time to perform than single qubit gates. This raised the question as to whether this additional time might be responsible for the increased qubit to qubit interactions seen in the two qubit protocol. Some further experiments were conducted to test this. Rather than activating multiple two qubit resonators simultaneously only two qubits were run in the two qubit protocol, the other 12 qubits being operated in `single qubit' mode. This was done for all physically possible qubit to qubit pairs. In addition the single qubit protocol was repeated, but this time a wait equivalent to the activation of two-qubit gate was imposed, only one of the 14 qubits being driven the others idle. (Under the API for the device at the time, it was necessary to drive at least one of the qubits in order to ensure the correct wait time was observed.)

\begin{figure*}
	\centering
	\begin{tikzpicture}
	\node at (-10,0) {\includegraphics[width=0.7\textwidth]{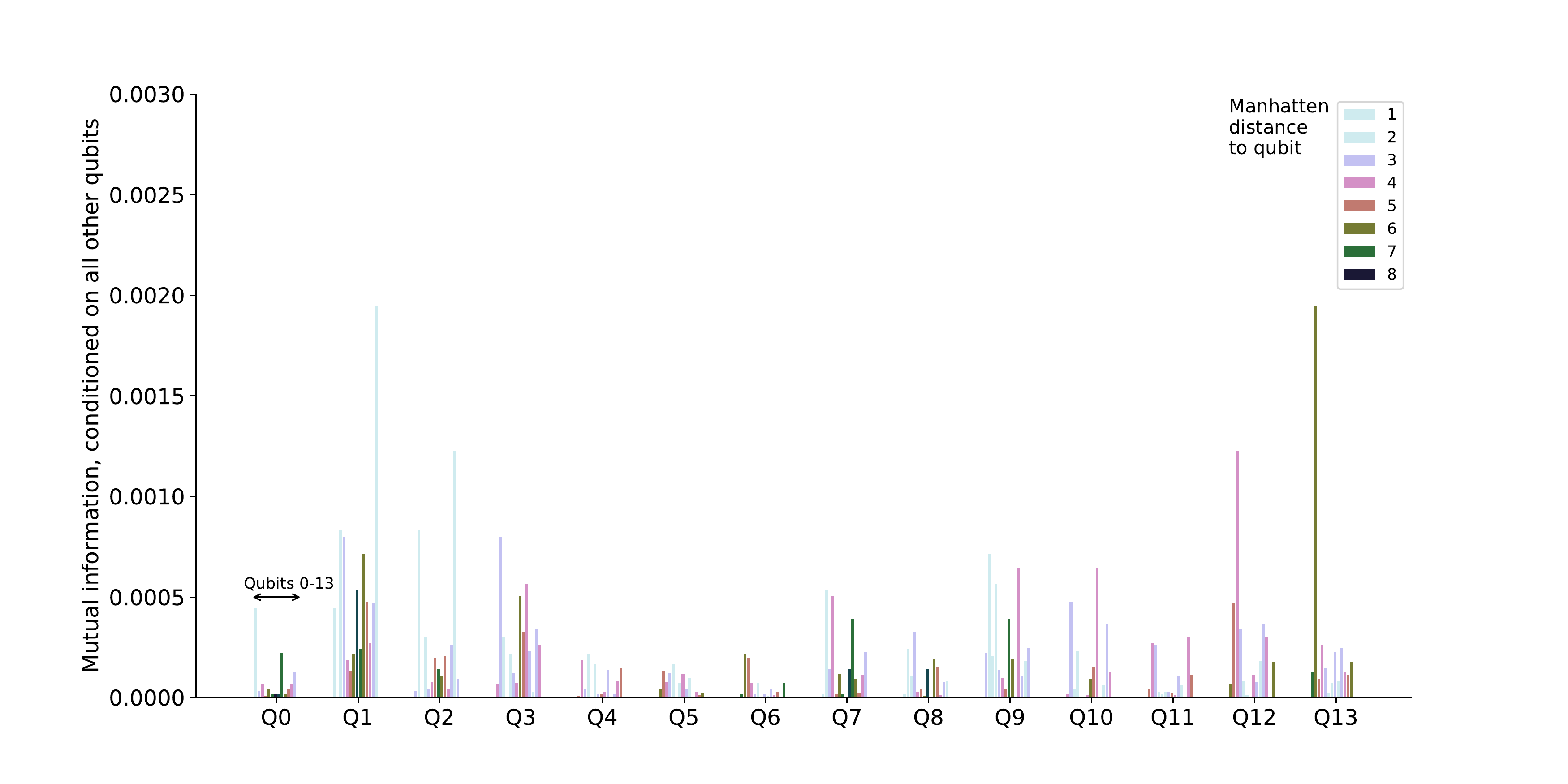}};
	\node at (-17,0) {(a)};
	\node at (-10,2) {Single qubit protocol};
	\node at (-10,-6) {\includegraphics[width=0.7\textwidth]{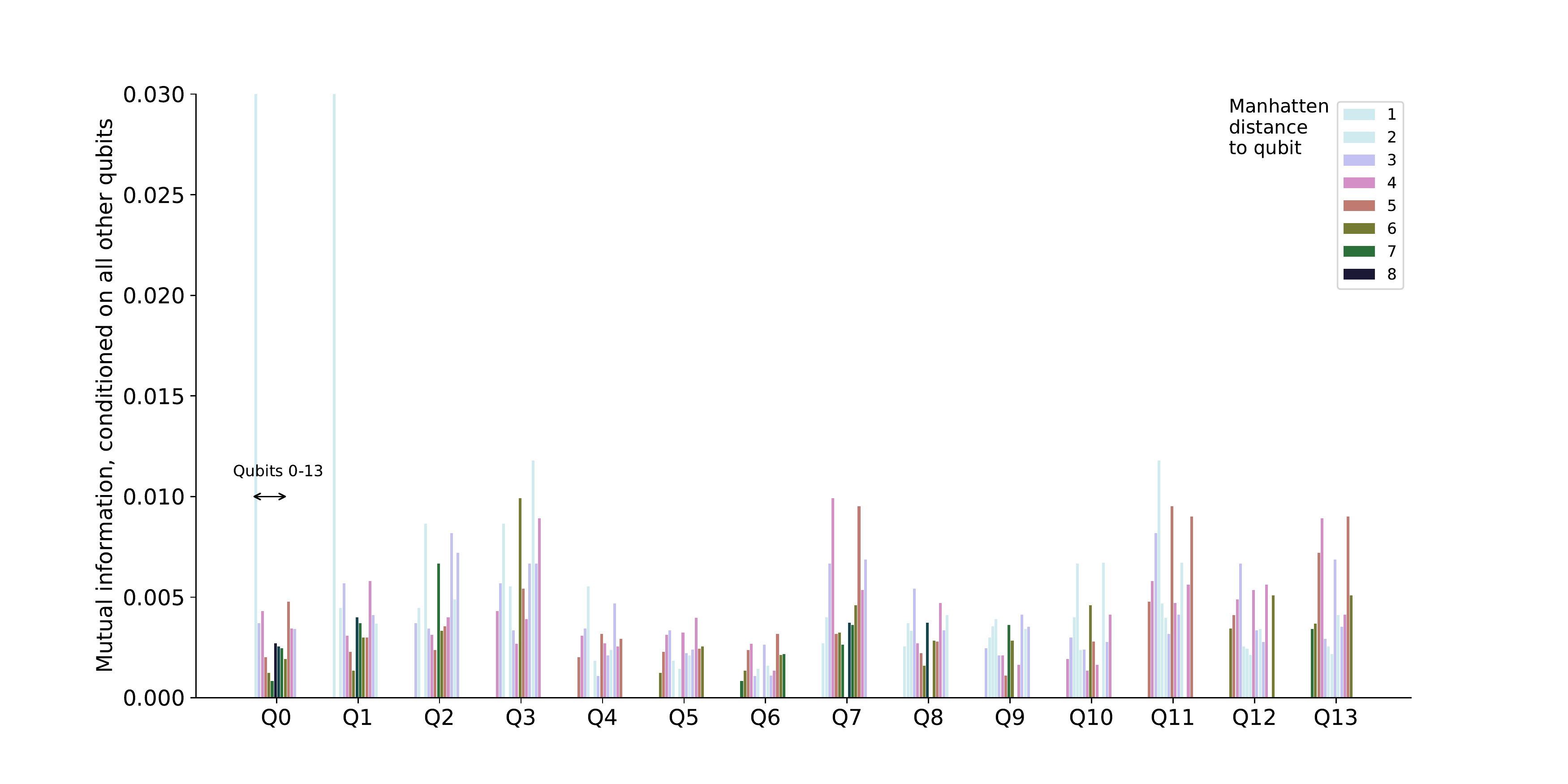}};
	\node at (-17,-6) {(b)};
	\node at (-10,-4) {Twin qubit protocol between Q0 and Q1};
	\node at (-10,-12) {\includegraphics[width=0.7\textwidth]{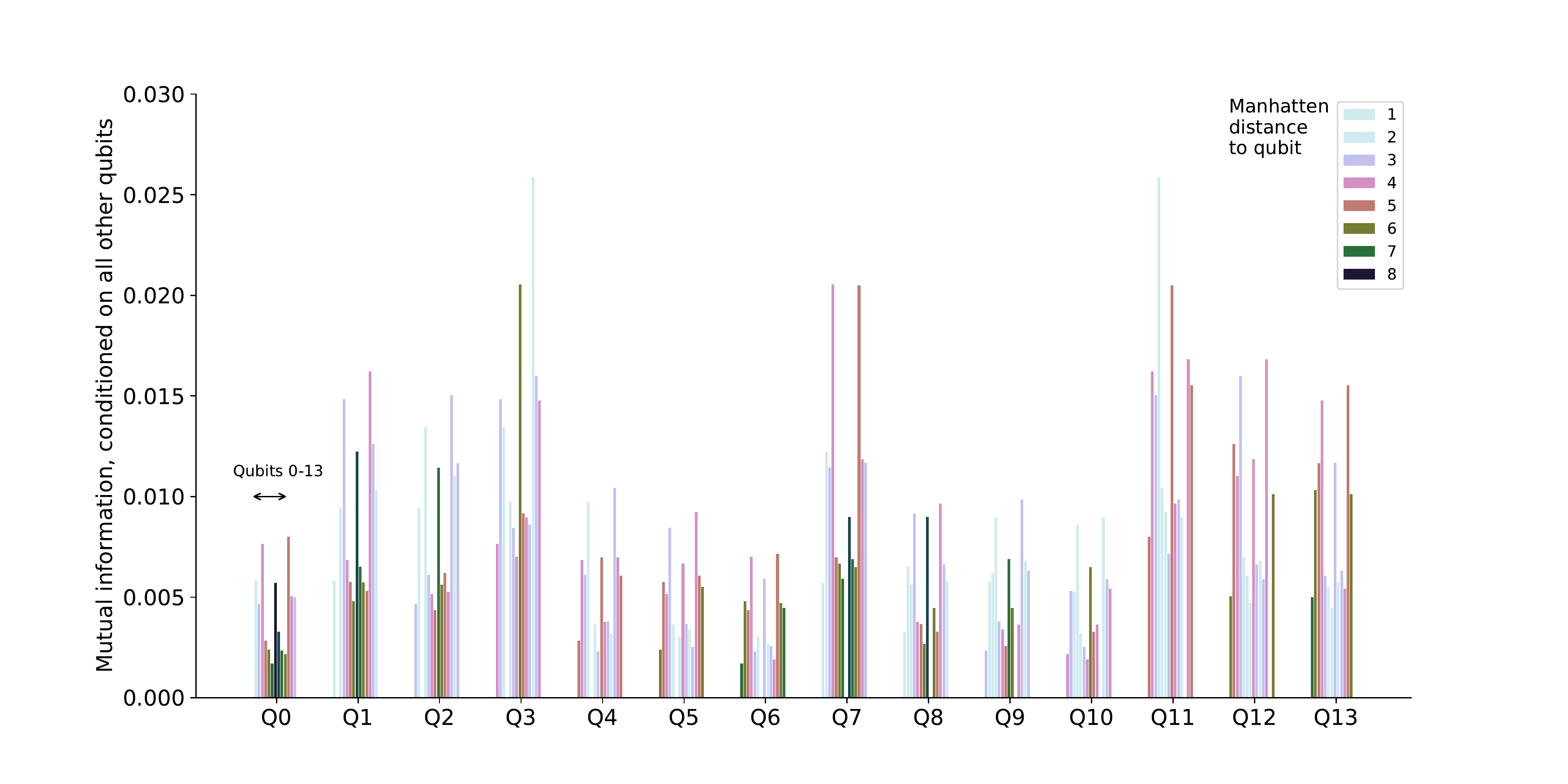}};
	\node at (-10,-10) {Single qubit protocol, enforced wait};
	
	\node at (-17,-12) {(c)};
	\end{tikzpicture}
	\caption{A series of graphs showing the conditional mutual information between each of the qubits (conditioned on all the others) in a variety of situations. For each qubit, the conditional mutual information between that qubit and each of the other qubits are plotted in a bar chart. The color of the bar represents the distance between that qubit and the corresponding qubit (based on physical Manhatten distance of the device). The length of the bar represents the value of conditional mutual infromation between the qubit represented by the x-axis and the qubit represented by the position of the bars. The bars are sequentially laid out as qubit 0 to qubit 13 (as shown for qubit 0 in the charts). Chart (a) shows the mutual conditional information of the averaged noise where the device is operated in single-qubit protocol. Note the scale of the y-axis. It is an order of magnitude smaller than the next two charts. (b) shows the device where a two-qubit twirl was conducted between qubits 0 and 1 (thus, the large lines for these two qubits). The remaining qubits were driven with single qubit Clifford twirls. There is approximately 10 times more `mutual information' between all the qubits in question compared to figure (a). Finally chart (c) shows the averaged noise where the qubits are operated in single qubit mode, but all of them (other than qubit 0) are forced to idle for a length of time roughly equal to that required to perform a two qubit gate. As is clear the mutual information (i.e. cross-talk) between the qubits is of a similar order of magnitude to (b).}\label{fig:conditional}
\end{figure*}
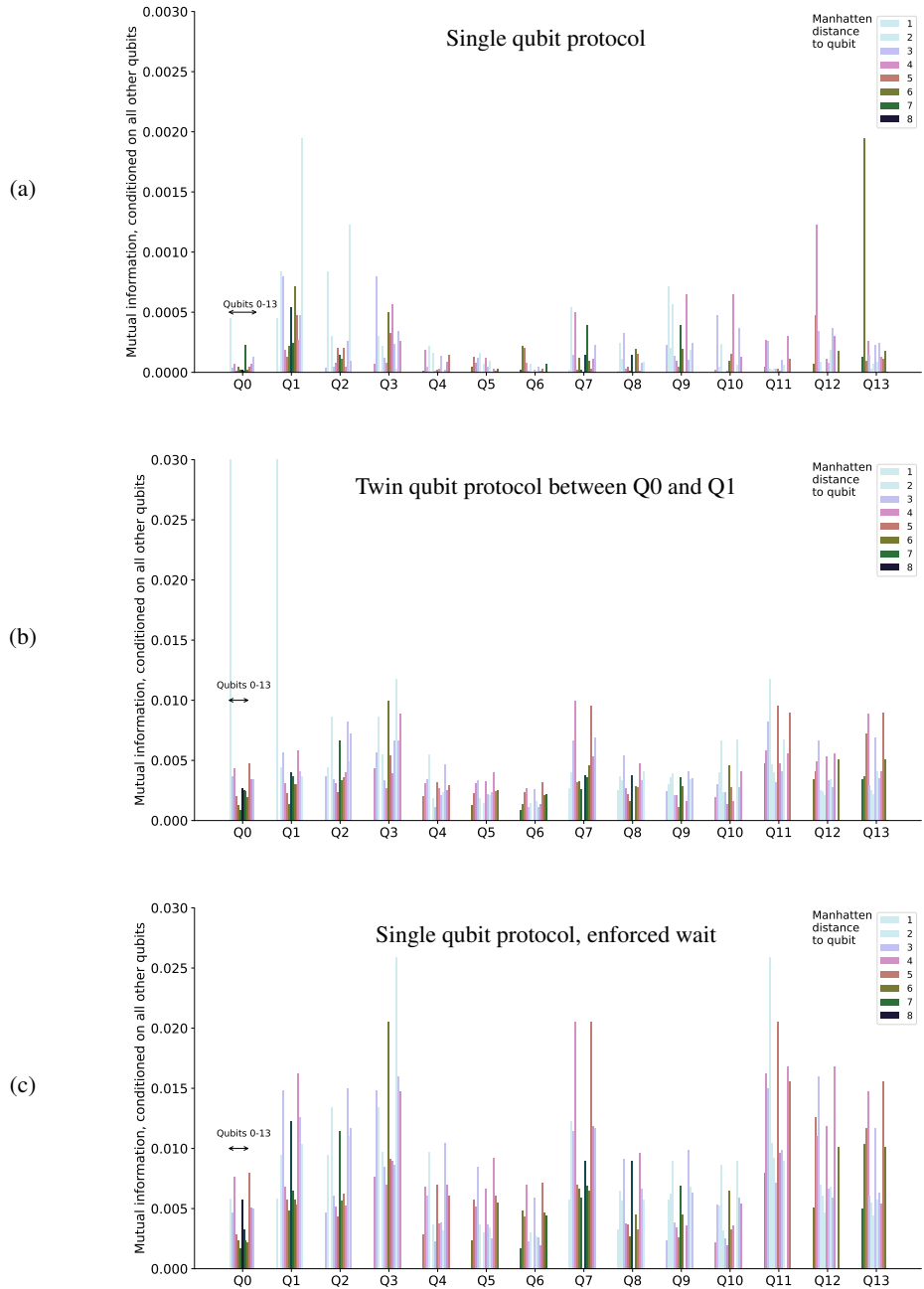

We used conditional mutual information (see Methods) to analyse these results. We were interested in whether given knowledge of all other qubits, how much information still existed between two qubits in question. If the qubits were independent this would be 0, with a maximum of 1 (i.e. completely dependent). In \cref{fig:conditional} we show the results for the full single qubit twirl, the experiment that performed the two qubit twirl between qubits 0 and 1, and the experiment that imposed a wait time. As can be seen it would appear that most of the long-range correlations we see arise during the two qubit protocol also arise if we impose a wait time similar to the length of time required to perform the two qubit gates. Whilst our access and knowledge of the machine is limited and therefore we do not wish to speculate as to causes, this is another example of the type of information that is revealed from knowledge of the locally averaged global probability distribution.

\section{Additional information relating to the IBM Quantum Experience}

As discussed in the main text there were two main experimental runs. The first was the 1000 runs used for the single qubit protocol. This took place between the 15th February 2019 and 16th Febraury 2019. \Cref{fig:timing}(a) contains a  chart showing the distribution of each of these runs (each run consisted of 11 sequences with 1024 shots per sequence). The two large gaps in the chart (at approximately 8am on each day) and the gate at 5pm were probably re-calibrations, but unfortunately we do not have specific data to confirm this.

Other gaps would have been queuing issues  or down time while we waited for `credits' to be regenerated.

The second experiment used batch jobs (since the run lengths were shorter) -  each batch consisted of six runs (being two of each of the different qubit configurations). At the time the IBM Q experience would occasionally `swallow' jobs which were permanently shown as running, and so we only managed to retrieve 973 of the jobs, equating to the 1946 runs for each configuration of two qubit gates (each run consisting of 11 sequences, with 1024 shots per sequence). The timing of the runs is shown in \cref{fig:timing}(b)

We were monitoring the results reported by the backend at the time of the second experiment and they were reported as updating the backend statistics at the following times, of which only one falls within the timestamps of the pickles saved:

\begin{quote}
2019-03-12 08:30:46+00:00

2019-03-14 04:53:05+00:00

2019-03-17 07:01:33+00:00

2019-03-18 06:48:41+00:00
\end{quote}

Because we were saving the backend status reports at the time of the two qubit runs, we can reproduce the reported T1s and T2s in \cref{table:t1s}.

\begin{table}[ht]
\begin{tabular}{c|l l|l l}
\textbf{Qubit} &    \textbf{2019-03-14} &		 &   \textbf{2019-03-17}\\
&T1&T2&T1&T2\\
\hline
 
0 & 4.218e-05& 1.912e-05& 7.160e-05  & 2.104e-05 \\
1 & 3.873e-05& 3.971e-05& 6.143e-05  & 1.198e-04 \\
2 & 6.679e-05& 1.107e-04& 5.560e-05  & 9.419e-05 \\
3 & 6.453e-05& 4.213e-05& 7.830e-05  & 6.618e-05 \\
4 & 5.126e-05& 3.140e-05& 8.281e-06  & 1.070e-05 \\
5 & 2.305e-05& 4.288e-05& 2.521e-05  & 4.281e-05 \\
6 & 9.449e-05& 5.949e-05& 9.449e-05  & 2.870e-05 \\
7 & 4.737e-05& 6.835e-05& 4.737e-05  & 3.122e-05 \\
8 & 5.352e-05& 6.277e-05& 5.689e-05  & 9.027e-05 \\
9 & 4.725e-05& 6.731e-05& 4.287e-05  & 7.272e-05 \\
10 & 6.005e-05& 6.944e-05& 5.896e-05  & 7.046e-05 \\
11 & 5.787e-05& 7.954e-05& 6.796e-05  & 1.235e-04 \\
12 & 6.780e-05& 9.217e-05& 8.905e-05  & 1.547e-04 \\
13 & 2.364e-05& 4.347e-05& 1.972e-05  & 2.705e-05 \\
\end{tabular}\caption{Reported T1 and T2 times during the course of the experiment using the  two-qubit protocol. Only one change to the backend statistics occured during the course of the experiment.}\label{table:t1s}
\end{table}

Full backend information during the course of the second run is included with the information and data made available with this paper.

\end{document}